\documentclass[twocolumn,showkeys,preprintnumbers,floatfix,amsmath,a4paper,nofootinbib,amssymb]{revtex4}

\usepackage[latin1]{inputenc}
\usepackage{graphicx}
\usepackage{amsthm}

\usepackage{graphicx}
\usepackage{bm}
\usepackage{graphicx}

\begin{document}

\title{Stem cell lineage survival as a noisy competition for niche access}

\author{Bernat Corominas-Murtra$^1$, Colinda L.G.J. Scheele$^2$, 
Kasumi Kishi$^1$, Saskia I.J. Ellenbroek$^2$, Benjamin D. Simons$^{3,4,5}$, Jacco van Rheenen$^2$, Edouard Hannezo$^1$}

\affiliation{$^1$Institute for Science and Technology Austria. Am Campus 1, A-3400 Klosterneuburg, Austria\\
$^2$Division of Molecular Pathology, Oncode Institute, The Netherlands Cancer Institute, Plesmanlaan 121, 1066 CX, Amsterdam, The Netherlands\\
$^3$The Wellcome Trust/Cancer Research UK Gurdon Institute, University of Cambridge, Cambridge CB2 1QN, UK\\
$^4$Department of Applied Mathematics and Theoretical Physics, Centre for Mathematical Sciences, University of Cambridge, Wilberforce Road, Cambridge CB3 0WA, UK\\
$^5$The Wellcome Trust/Medical Research Council Stem Cell Institute, University of Cambridge, Cambridge CB2 1QN, UK}

\thanks{To whom correspondence should be addressed. E-mail: bernat.corominas-murtra@ist.ac.at, j.v.rheenen@nki.nl, edouard.hannezo@ist.ac.at\\
$^*$Author contributions: B.C.-M., B.D.S, J.v.R and E. H. designed research; B.C.-M., C.L.G.J.S, K.K, S. I.J. E., and EH performed research and analyzed data; B.C-M and E.H. wrote the paper with input from C.L.G.J.S, B.D.S and J.v.R.\\
$^*$B.C.-M. and C.L.G.L.S. contributed equally to this work.}


\begin{abstract}
Understanding to what extent stem cell potential is a cell-intrinsic property, or an emergent behavior coming from global tissue dynamics and geometry, is a key outstanding question of systems and stem cell biology. Here, we propose a theory of stem cell dynamics as a stochastic competition for access to a spatially-localized niche, giving rise to a stochastic conveyor-belt model. Cell divisions produce a steady cellular stream which advects cells away from the niche, while random rearrangements enable cells away from the niche to be favourably repositioned. Importantly, even when assuming that all cells in a tissue are molecularly equivalent, we predict a common ("universal") functional dependence of the long-term clonal survival probability on distance from the niche, as well as the emergence of a well-defined number of functional stem cells, dependent only on the rate of random movements vs. mitosis-driven advection. We test the predictions of this theory on datasets on pubertal mammary gland tips, embryonic kidney tips as well homeostatic intestinal crypt. 
Importantly, we find good agreement for the predicted functional dependency of the competition as a function of position, and thus functional stem cell number in each organ. This argues for a key role of positional fluctuations in dictating stem cell number and dynamics, and we discuss the applicability of this theory to other settings.
\end{abstract}


\maketitle

Many biological tissues are renewed via small numbers of stem cells, which divide to produce a steady stream of differentiated cells and balance homeostatic cell loss. Although novel experimental approaches in the past decade have produced key insights into the number, identity, and (often stochastic) dynamics of stem cells in multiple organs, an outstanding question remains as to whether stem cell potential is a cell-intrinsic, "inherited" property, or rather an extrinsic, context-dependent state emerging from the collective dynamics of a tissue and cues from local "niches", or microenvironments \cite{Wang:2010, Simons:2011, Blanpain:2013,  Watt:2013, Blanpain:2014, Plaks:2015, Moris:2016, Yang:2017}. Although recent experiments have provided evidence for the latter in settings such as the growing mammary gland \cite{Scheele:2017}, adult interfollicular epidermis \cite{Rompolas:2013, Rompolas:2016}, spermatogenesis \cite{Kitadate:2019} or the intestinal epithelium \cite{Ritsma:2014}, a more global theoretical framework allowing to quantitatively interpret these findings is still lacking.

The case of the intestinal crypt serves as a paradigmatic example of the dynamics of tissue renewal, and is one of the fastest in mammals \cite{Ritsma:2014}. The intestinal crypt consists of a small invagination in the intestine where the epithelial cells populating the intestinal walls are constantly produced. The very bottom of the crypt hosts a small number of proliferative, Lgr5+ stem cells, \cite{Barker:2007} that divide and push the cells located above them to the transit amplification (TA) region, where cells lose self-renewal potential. Cells are eventually shed in the villus a few days later, constituting a permanent "conveyor-belt" dynamics. Lineage tracing approaches, which irreversibly label a cell and its progeny \cite{Blanpain:2013}, have been used to ask which cell type will give rise to lineages that renew the whole tissue and have revealed that all Lgr5+ cells can stochastically compete in an equipotent manner on the long term \cite{Snippert:2010,Lopez-Garcia:2010,Snippert:2011, Klein:2011}, but still display positional-dependent short term biases for survival \cite{Ritsma:2014}. Interestingly, similar conclusions have been reached in pubertal mammary gland development \cite{Scheele:2017}, where branching morphogenesis occurs through the proliferation of the cells in the terminal end buds of the ducts \cite{hannezo2017unifying}, the region where the mammary stem cells (MaSCs) reside \cite{Scheele:2017, Visvader:2014}. In both cases, intravital imaging revealed random cellular motions enabling cells to move against the cellular flow/drift defined by the conveyor belt dynamics. Moreover, in the intestine, tissue damage, or genetic ablation of all Lgr5+ stem cells, caused Lgr5- cells to recolonize the crypts and re-express Lgr5+ to function as stem cells \cite{Ritsma:2014}, arguing for extensive reversibility and flexibility in the system \cite{vanEs:2012}. In addition, Lgr5- and Lgr5+ cells of the fetal gut were also shown to nearly equally contribute to intestinal morphogenesis \cite{Guiu:2019}. Altogether, this supports proposals that the definition of stem cell potential should evolve to emphasize, instead of molecular markers, 
the functional ability of cells to renew over the long-term \cite{Krieger:2015, Post:2019}. 

However, this new definition raises a number of outstanding conceptual problems: What then defines the number of functional stem cells in a tissue? How can short-term biases be reconciled with long-term equipotency? Is there a sharp distinction between stem and non-stem cells, or is there instead a continuum of stem cell potential together with flexible transition between states? Qualitatively, it is clear that fluctuations and positional exchanges are needed to prevent a single cell in the most favourable position to be the unique "functional" stem cell (defined as cells whose lineage colonizes a tissue compartment on the long-term). Incorporating these features in a dynamical model of stem cell growth and replacement, able to make predictions e.g., on the probability of lineage perpetuation, would represent an important step towards the understanding of how stem cells operate in the process of tissue growth and renewal.

\begin{figure*}[ht]
\includegraphics[width= 17.7cm]{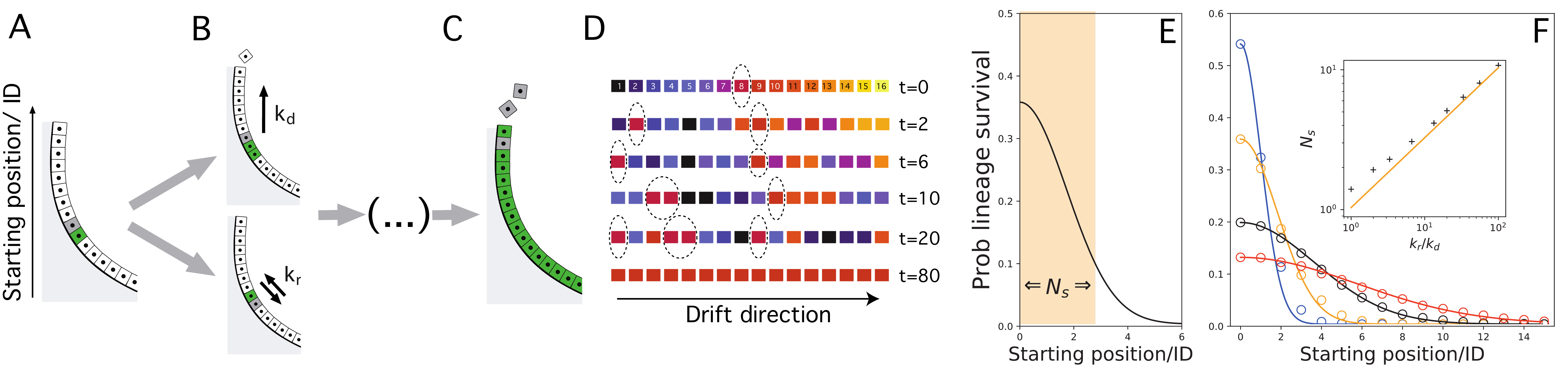}
\caption{Stochastic conveyor belt as a paradigm for stem cell renewal. A/ A cell in the epithelial wall of the crypt can B/ duplicate at rate $k_d$ pushing the upper cells up, creating a conveyor-belt mechanism; or switch its position randomly at rate $k_r$, introducing a stochastic or noisy ingredient in the dynamics. C/ At longer time scales, the lineage of a single starting cell colonizes the whole system. D/ Example of SCB dynamics. At $t=0$ we have $N=16$ lineages in the system, depicted with different colors and at starting positions $1, . . .,16$ respectively. In time, lineages are progressively eliminated, but stochastic cell rearrangements makes it possible for a lineage far from the origin (starting position $n=8$ in red and highlighted with a dashed circle) to win the competition.
E/ Probability that a given lineage colonizes the entire system as a function of initial position of its mother cell, decaying as a Gaussian of width $\sqrt{k_r/k_d}$, see text for details.  The width of this distributions defines a functional stem cell region ($N_s$ cells, highlighted in orange, plotted for $k_r/k_d=3$).  F/ Numerical simulations of the 1-dimensional SCB dynamics. We compute the  long term survival probability $p(c_n)$ as a function of initial starting position $n=0,1,2,...$, with respect to the base of the system for several values of $k_r/k_d$ ($1$, $3.3$,  $13.3$ and  $33$ in resp. blue, orange, black and red). Dots show the outcome of the simulations and lines show the analytical prediction $p(c_n)\sim {\rm exp}\{-\frac{k_d}{2k_r}n^2\}$, as shown in equation (\ref{eq:p(c_k)}).
Inset shows plot of best fit for the variance of the numerical distributions (black crosses) against the analytical model prediction $\sim\sqrt{k_r/k_d}$ (orange solid line). }
\label{fig:Schema}
\end{figure*}

In this paper we develop a reaction-diffusion formalism for stem cell renewal in the presence of noise and local niches, taking into account local tissue geometry as well as cell division and random cell movements (Fig. \ref{fig:Schema}a--c). Importantly, within this purely extrinsic and dynamical approach, which does not need to posit any intrinsic "stem cell identity", a well-defined number of functional stem cells emerges, which only depends on the geometry and a balance between the noisiness of cell movements and division rates advecting cells away from niche regions. This model also predicts that stem cell potential should decay continuously as a function of distance from the niche, with a "universal" Gaussian functional dependence. We test this prediction against published live-imaging datasets for the homeostatic intestinal crypt \cite{Ritsma:2014} and during the branching of embryonic kidney explants \cite{Riccio:2016}, and find a good quantitative agreement for the full survival probability of cells depending on their initial position relative to the niche.  Furthermore, we use our theoretical results to extract the amplitude of the random positional fluctuations in the developing mammary gland using static lineage tracking experiments \cite{Scheele:2017}. This enables us to predict the number of functional stem cells for this system, finding values consistent with previously reported estimates.

\section*{Dynamics of tissue renewal and development}

To develop the model, we first consider the simplest situation of a one-dimensional column of cells, with a rigid boundary condition at the base (mimicking, for instance, the bottom of the crypt), so that each cell division produces a pushing force upwards transmitted to the cells above (or in the case of growing mammary gland or kidney, driving ductal elongation). This model is motivated by its simplicity, as it is able to qualitatively derive the essential traits of the complex dynamics studied here. As we shall see, further refinements, aimed at making predictions for real systems, consider  more realistic geometries. From this simple dynamics, we define the number of functional stem cells as the typical number of cells that have a non-negligible probability to produce long-term progenies (without "losing" the competition against other cells). If the dynamics was fully devoid of noise (a simple conveyor belt) and all cell divisions were symmetric, then one of the bottom-most cells would always win the competition. In the case of a 1-dimensional array of cells, this problem is trivial. If one considers a cylindric geometry, there would be a single row of functional stem cells, which is the limiting case of the model described in Ref. \cite{Lopez-Garcia:2010} of symmetric and stochastic 1-dimensional, neutral competition along a ring of equipotent cells. 
However, live-imaging studies shows that, in multiple settings including mammary gland  \cite{Scheele:2017}, kidney morphogenesis \cite{Riccio:2016, Packard:2013} and intestinal crypts \cite{Ritsma:2014}, there is widespread rearrangement of cells through stochastic cell movements.
Intuitively, such rearrangements are expected to increase the number of "functional" stem cells, as re-arrangements allow cells away from the niche to relocate to favourable positions, and would thus provide a biophysical mechanism for setting the number of stem cells assumed in models such 
such as that developed in Ref. \cite{Lopez-Garcia:2010}.

The simplest abstraction of the system is a 1-dimensional column of $N$ cells. Each cell divides at constant rate $k_d$. In 1D, we assume a rigid boundary at the bottom so that cell proliferation generates a net flow of cells along the positive axis, i.e. advection away from the niche. In addition, the position of the cells can fluctuate stochastically at rate $k_r$ (either via local cell-cell rearrangements, or more global movements of cells relative to the niche, see sections  1A and  4 of the appendixfor details), allowing cells far away from the niche to reposition despite the overall flow.

At $t=0$, each cell is characterised by its starting position $n$ (distance from the niche), and will give rise in time to a lineage denoted $c_n$, which can span the entire tissue. However, as soon as a cell reaches the position $N$, it disappears from the system, resulting after a sufficiently large time period in a single surviving lineage. This competitive dynamics can be metaphorically understood as a conveyor belt with random fluctuations in the cell positions, sketched in Fig. \ref{fig:Schema}a--c. This is why we call it {\em Stochastic Conveyor Belt} (SCB) dynamics, and use it to model tissue renewal (e.g. intestinal crypt homeostasis) or organ growth (e.g. kidney and mammary gland morphogenesis). The only difference between these two general cases is a change of reference frame (see section  1 and Fig. S1 of the appendix). In Fig. \ref{fig:Schema}d, we show an example of a typical run of the simulated SCB dynamics in 1 dimension, until monoclonality is achieved (see  section 5A of the appendix for details).

To make quantitative predictions from the dynamics outlined above, we start by following the prevalence of a single lineage. Here, the action of the other lineages can be imposed as an average drift force that depends on the position of each cell of the lineage we follow.  The equation accounting for the time evolution of the prevalence of lineage $c_n$, to be referred to as $\rho_n(z,t)$, in the continuum limit is:
\begin{equation}
\frac{\partial \rho_n}{\partial t}=-k_d\frac{\partial }{\partial z}(z\rho_n)+\frac{k_r}{2}\frac{\partial^2 \rho_n}{\partial z^2}+k_d\rho_n\quad.
\label{eq:FisherKPP}
\end{equation}
We refer to this reaction-diffusion equation \cite{Britton:1986, Grindrod:1996} as the SCB equations (see appendix, section 1B for details). The first term on the right hand-side is a drift term, accounting for the average push up movement at position $z$ due to random cellular proliferation at rate $k_d$ at lower levels, $\sim k_d z$. The second term is a diffusive term \cite{Gardiner:1983, VanKampen:2007} accounting for the random reallocations of cells, occurring at rate $k_r$. The third term is a proliferative term, accounting for the exponential proliferation of each cell of the lineage under study, at rate $k_d$. 

Considering initial conditions $t_0=0$, $\rho_n(z,0)$ a Gaussian centered on $n$ with $\sigma^2=1/2$ (a density representing a single cell at position $n$) and natural boundary conditions, 
the solution of equation (\ref{eq:FisherKPP}) can be approximated by (see appendix, section 1B, for details):
\begin{equation}
\rho_n(z,t)\approx \sqrt{\frac{k_d}{2\pi k_r}}{\rm exp}\left\{-\frac{k_d}{2 k_r}\left(\frac{z-ne^{k_d t}}{e^{k_d t}}\right)^2\right\}\quad.
\label{eq:rhoexplicit}
\end{equation}

Next, we sought to relate this lineage prevalence to the experimentally relevant quantity of long-term lineage survival, in other words, {\em how likely is it for a cell starting at a given position $n$ to take over the entire crypt?}. Although lineage fixation is a concept that only makes sense in the discrete lineage,  we observed that lineage prevalence converges asymptotically towards a simple scaling form $\rho_n(\infty)$:
\begin{equation}
\rho_n(\infty)\equiv\lim_{t\to \infty}\rho_n(z,t)\quad,
\label{eq:rhoinf=}
\end{equation}
which is a constant that does not depend on position $z$ or time $t$, but only on the starting position of the lineage. This argues that, on the long-term, lineages starting at different positions $n$ and $n'$ have well-defined relative prevalence, leading to the natural assumption that the long term lineage survival probability of lineage $c_n$ is proportional to this asymptotic lineage prevalence. 
This means that the probability of lineage survival, $p(c_n)$, can be expressed as:
\begin{equation}
p(c_n)\approx\frac{\rho_n(\infty)}{\sum_j\rho_j(\infty)}\propto  {\rm exp}\left\{-\frac{k_d}{2k_r}n^2\right\}\quad.
\label{eq:p(c_k)}
\end{equation}
The above equation, which is a central result of the study, defines the probability that a cell starting at position $n$ will "win the competition" and colonize the whole one-dimensional system (see appendix, section 1 for details).

In spite of the approximations outlined above, stochastic numerical simulations of the model system show excellent agreement with equation (\ref{eq:p(c_k)})(Fig. \ref{fig:Schema}e,f). We also note that although we have assumed here that positional rearrangements occur between two cells, more complex sources of positional noise $k_r$ can be considered (which can be mechanistically dependent or independent on $k_d$), and lead to the same qualitative results. These include, for instance, post-mitotic dispersal, as seen during the branching morphogenesis of the kidney uteric bud \cite{Packard:2013} and where daughter cells can travel long distances outside the epithelium post-division, or correlated "tectonic" movements of the epithelium, where cells could collectively reposition relative to the niche, as proposed during mammary or gut morphogenesis \cite{Scheele:2017, Guiu:2019} (see appendix, section  4 for details).

\section*{Functional stem cell numbers and dynamics in the stochastic conveyor belt}

The prediction for the probability of long-term lineage survival under the SCB dynamics is surprisingly simple, decaying as a Gaussian distribution as a function of position away the niche, with a length scale that is simply the amplitude of the stochastic fluctuations divided by the proliferation rate, $\sim\sqrt{k_r/k_d}$ (see Eq. (\ref{eq:p(c_k)})). Intuitively, cells close to the origin have the highest chance to win and survive, whereas this probability drops abruptly for cells starting the competition further away, i.e. around $N_s$ cell diameters away from the base, with:
\begin{equation}
N_s^{1D} =  1+\sqrt{\frac{k_r}{k_d}}\quad.
\label{eq:NsTrivial}
\end{equation}
Note that the first term satisfies the boundary condition that, in the case $k_r=0$, the system has a single functional stem cell (located at the base) in 1 dimension. Eq. (\ref{eq:NsTrivial}) thus implies that multiple rows of cells possess long-term self-renewal potential (as assessed for example, in a lineage tracing assay), emerging through their collective dynamics, and with a number that depends only on the ratio of the division to rearrangement rates (resp. $k_d$ and $k_r$). Although Eq. (\ref{eq:NsTrivial}) is the outcome of a 1-dimensional approximation, we show that it holds and can be generalized in more complex geometries  (see section  3 of the appendix). In particular, in a cylindrical 2-dimensional geometry, we show the functional stem cell number would simply be the same number $N_s^{1D}$ of cell rows (arising from the stochastic conveyor belt dynamics) multiplied by the number of cells per row (fixed by the geometry of the tissue). Moreover, the above result can be generalized, giving an estimate of $N_s$ for general geometries (see Eq. (S26) of the appendix, where we give the general expression for $N_s$ in arbitrary organ geometries). This general result will be at the basis of the forthcoming sections, when dealing to more realistic geometries to explore the dynamics of the organs under study. Importantly, our framework generalizes the work of Ref. \cite{Lopez-Garcia:2010}, as we do not fix the stem cell number $N_s$ explicitly, which rather emerges from an interplay between geometry and SCB dynamics, together with the competitive dynamics being qualitatively different in the flow direction (see section 1 of the appendix).

We now turn to experimental data to test whether the proposed dynamics can help predict the number of functional stem cells in several organs, as well as the evolution of the survival probability with starting position of a clone. Although the division rate $k_d$ is well-known in most systems considered, the stochastic movement rate $k_r$ is harder to estimate, and can potentially vary widely, from rather small in intestinal crypts \cite{Ritsma:2014}, to large in mammary and kidney tips, with extensive clonal fragmentation and random cell movements \cite{Scheele:2017, Riccio:2016}. 

\subsection*{Predictions on clonal dynamics and survival}
\begin{figure}
\begin{center}
\includegraphics[width= 9cm]{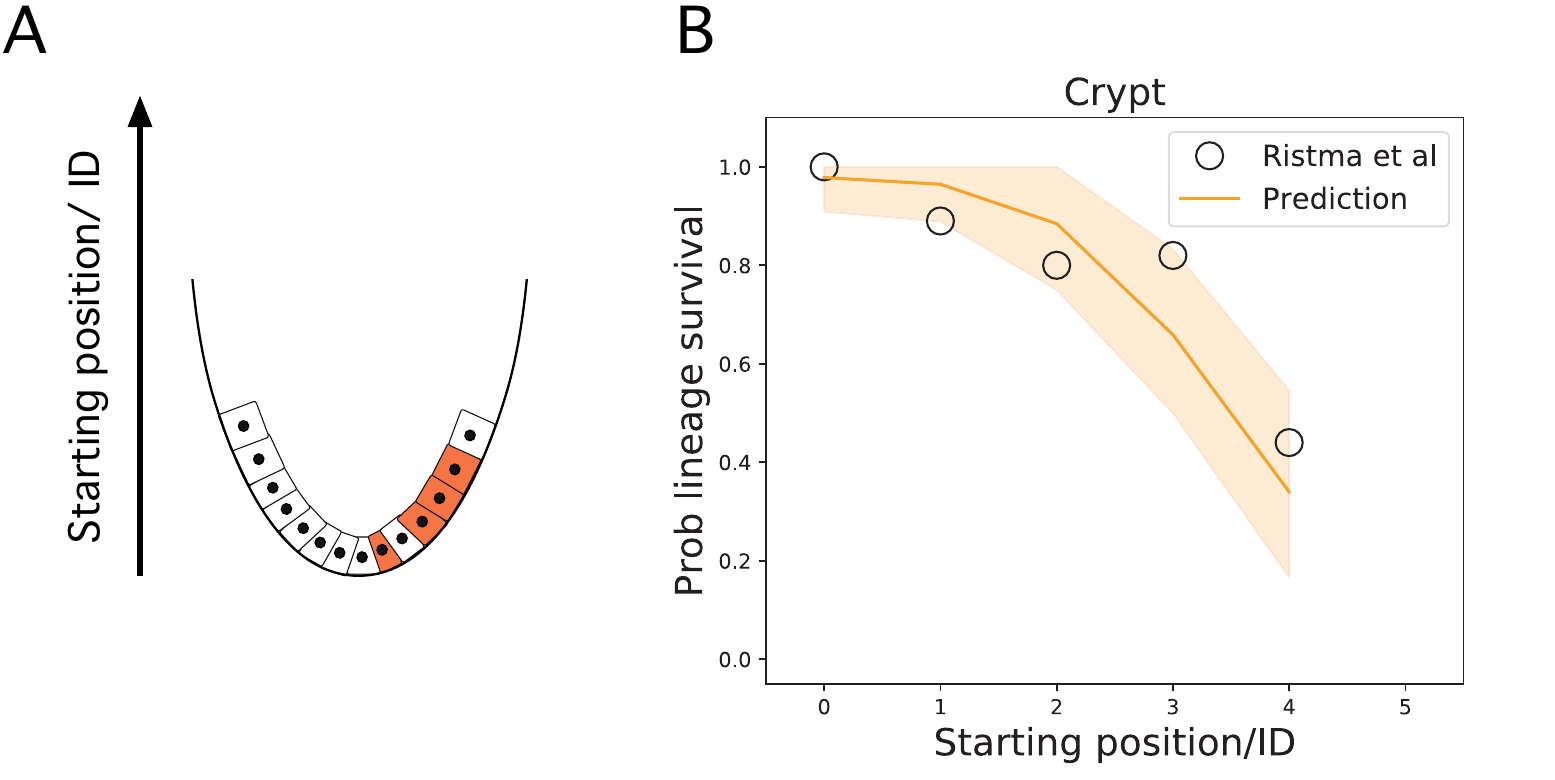}
\caption{ A/ Schema of the self-renewal of the crypt epithelia, showing the origin of the coordinate system at the bottom of the system. 
B/ The probability that a given lineage remains within the system as a function of the starting position after a time lapse against the predictions of the conveyor belt dynamics for the crypt. 
Data corresponding to the probability that a lineage remains in the system for the small intestinal crypt, reported in \cite{Ritsma:2014} depending on its starting position. The orange line represents the prediction of the stochastic conveyor belt dynamics, fitting well the data for $k_r/k_d \approx 1$. Shaded areas represent the confidence interval (1S.D.) of the prediction. }
\label{fig:RealSurvival}
\end{center}
\end{figure}

Intravital live-imaging provides an ideal platform to test the model, as it provides both knowledge of the starting position of a given cell as well as its clonal time evolution (whereas classical lineage tracing relies on clonal ensembles obtained from fixed samples). In small intestinal crypts, different Lgr5+ cells have been predicted to have very different lineage survival potential on the short-term, depending on their position within the stem cell niche, resulting in an effective number of stem cells smaller than the number of Lgr5+ cells \cite{Ritsma:2014, Kozar:2013}. We thus reanalyzed quantitatively this dataset by plotting the survival probability of a clone as a function of its starting position $n$ (Fig. \ref{fig:RealSurvival}) after a given time period assumed to be large enough for equation (\ref{eq:p(c_k)}) to hold. We then compared this to a 2-dimensional stochastic simulation of the model (see appendixsection 5 for details). Importantly, we found a good qualitative and quantitative agreement between model and data, with the survival probability decaying smoothly with the starting position (Fig. \ref{fig:RealSurvival}b). The only parameter here was $k_r/k_d \approx 1$, which fits well with short-term live imaging experiments and the idea of cell division promoting rearrangements \cite{Ritsma:2014}. 

To back these simulations with an analytical prediction on stem cell numbers, the details of tissue geometry must be taken into account (with the number of cells per row $i$ needing to be estimated, while the number of rows participating in the competition arising as an emergent property from the 1-dimensional model). A good approximation is based on that fact that the crypt can be abstracted as a hemispherical monolayer with radius $R$ (measured in units of cell diameter) coupled to a cylindrical region (Fig. S1, S3, S4 and section 3 of the appendix for details), so that one can get the number of stem cells, $N^{2D}_s$, as:
\begin{equation}
N^{2D}_s\approx 2\pi R^2\left[1-\cos\left\{ \frac{1}{R}\left(1+\sqrt{\frac{\pi}{2}\frac{k_r}{k_d}}\right)\right\}\right]\quad.
\label{eq:Ns}
\end{equation}
With $k_r/k_d \approx 1$ as above, and estimating $R \approx 2$ for the radius, our simple theory then predicts that the number of functional stem cells should be $N^{2D} \approx 11$, which agrees well with measurements of \cite{Ritsma:2014}, as well as inferred numbers from continuous clonal labelling experiments \cite{Kozar:2013}. This is expected, as our model reduces to the 1-dimensional ring model of Ref. \cite{Lopez-Garcia:2010} for low $k_r/k_d$.
 \begin{figure}
\begin{center}
\includegraphics[width= 9cm]{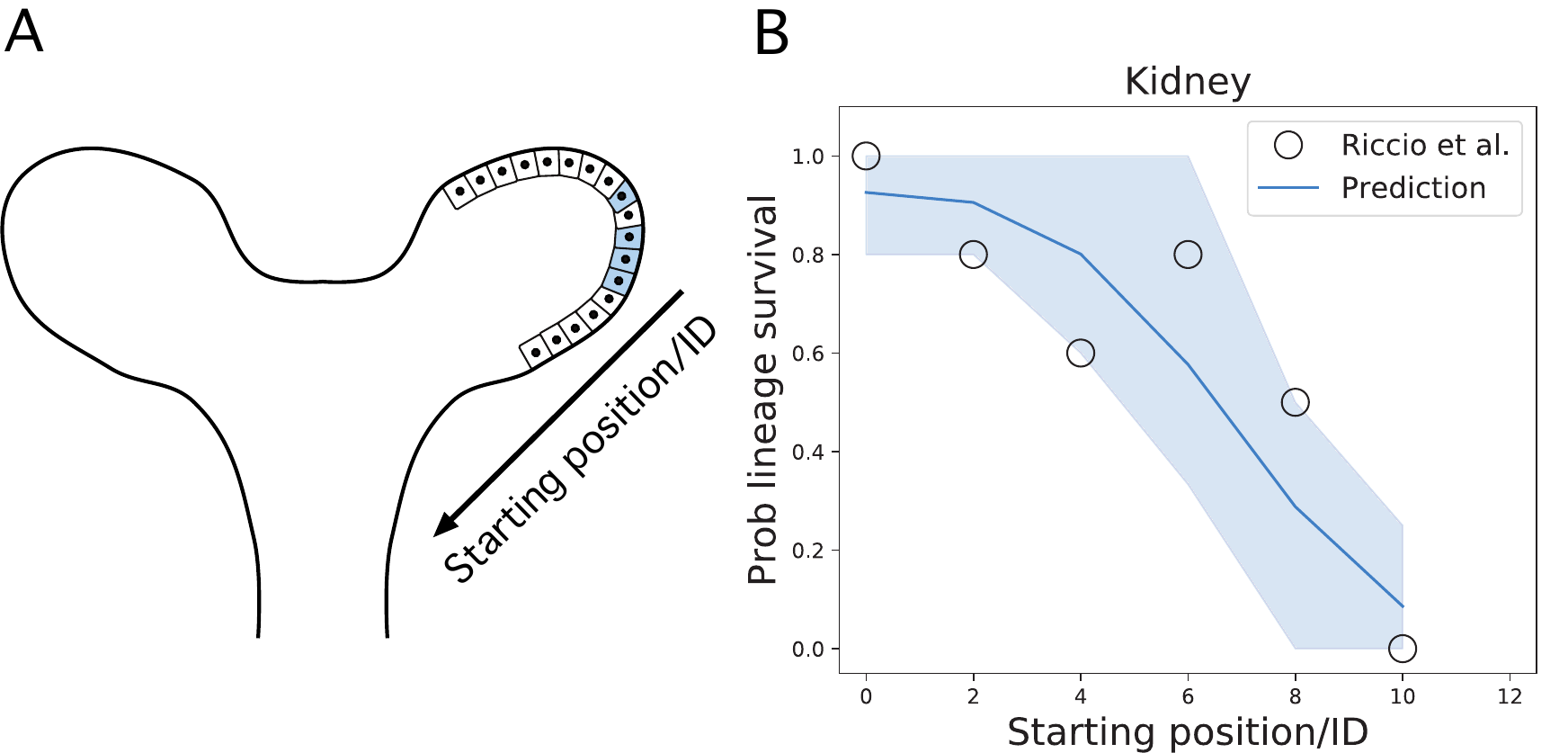}
\caption{ A/ Schema of the kidney tip during development. The conveyor-belt dynamics holds, the only difference is the reference frame: Whereas in the stem cell replacement model of the intestinal crypt the reference frame is the bottom of the gland, in the kidney and mammary gland, the reference frame is taken from the newly created ducts. B/ The probability that a given clonal remains within the system as a function of the starting position of the mother cell after a given time against the predictions of the conveyor belt model dynamics. Black circles represent real data points, obtained by counting the amount of cells of a given lineage remaining in the system (from Ref. \cite{Riccio:2016}). We observe that the distribution is much broader, fitting well to the theory for a ratio $k_r/k_d \approx 16$ in kidney, over an order of magnitude larger than in intestinal crypt.  Shaded area represents the confidence interval (1S.D.) of the prediction. }
\label{fig:RealSurvival_Kidney}
\end{center}
\end{figure}

We then sought to test the model further using a published dataset on embryonic kidney branching in explants \cite{Riccio:2016}. This has been recently noted to be a highly stochastic process, with neighbouring cells at the start of the tracing ending up either surviving long-term in tips or being expelled to ducts. Moreover, Ref. \cite{Riccio:2016} observed extensive random cell intercalations, in addition to the previously described mitotic dispersal \cite{Packard:2013}, where cells extrude from the epithelium post-division and reinsert at a distance of $d_c$ cell diameters away. Importantly, these processes can still be captured as an effective diffusion coefficient $k_r$ in our framework (see section 4 of the appendix for details). Specifically, knowing that the fluctuations may occur at each duplication, and that they imply a displacement up to $d_c \approx 2-4$ cell lengths, we can estimate that $k_r/k_d \approx d^2_c\quad$ at the minimum (i.e. discounting other fluctuations). Note that the conveyor belt dynamics applies exactly for tip elongation as in crypt: The only difference is that the reference frame from which the dynamics is observed changes (see section 1 and figure S1 of the appendix for details).

The above observation argues again that noise will play a key role in kidney tip cell dynamics. Strikingly, extracting from Ref. \cite{Riccio:2016} the probability of survival as a function of distance from the edge of a tip, we found that the 2-dimensional simulations of our model provided again an excellent prediction for the full probability distribution (Fig. \ref{fig:RealSurvival_Kidney}a,b), with cells much further away (compared to the intestinal crypt) having a non-negligible probability to go back and contribute. Again, the only fit parameter was the ratio $k_r/k_d = 16$, which agrees well with our estimate of the noise arising from mitotic dispersal. Taking into account the full 2-dimensional geometry as above, and estimating in this case a tip radius of $R=3-5$ cells, this predicts $N_s \approx 90\pm 10$, which could be tested in clonal lineage tracing experiments. 

These two examples show that the same model of SCB dynamics and its prediction of the master curve for the survival probability of clones can be used in different organs to understand their stem cell dynamics, and shows that ratios of relocation to advection $k_r/k_d$ can be widely different even in systems with similar division rates $k_d$.

\subsection*{Number of functional stem cells in the developing mammary gland}

\label{sec:Mam}
\begin{figure}[h]
\includegraphics[width= 8.9cm]{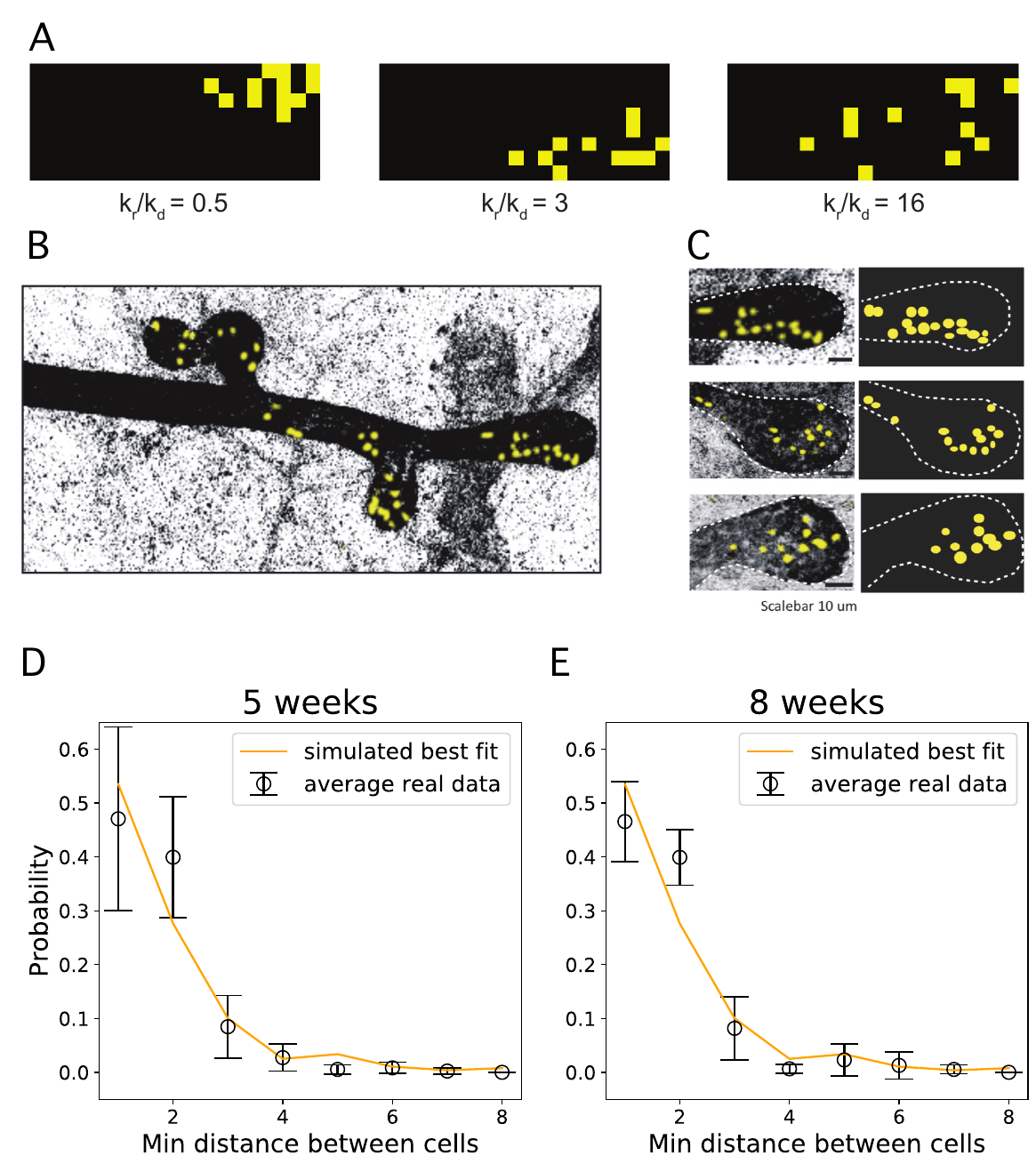}
\caption{A/ Inferring the relation $k_r/k_d$ from the clone dispersion using a simulation of the stochastic conveyor belt dynamics in 2 dimensions. The distribution of distances of the closest neighbours is highly sensitive to the relation $k_r/k_d$. Here we show numerical simulations of fragmentation under increasing (left to right) values of $k_r/k_d$.
B/ Growing tips of a developing mammary gland together with sparse lineage-tracing experiments, where a single lineage (yellow here, induced in 3w-old animals) can be observed. Clonal dispersion due to random cell rearrangements is observed. C/ Close-up of three different mammary tips (left) and corresponding reconstructions to extract relative cellular positions (right). The geometry of the end buds can be approximated by a hemispherical structure connected to a cylindrical one whose radius can be inferred to be around $2-5$ cell diameters. 
D-E/ Probability distributions of nearest distances between clonally-related cells in tips (resp. from 5-week and 8-week old mice). Black dots represent experimental data (basal and luminal cells have been treated together for this analysis, as they do not show different behaviour at the level of the dynamics). Orange lines are from 2-dimensional numerical simulations of the SCB model (see appendixtext for details) showing a good fit from $k_r/k_d\approx 3$ for both time points. Error bars represent mean and SD.}
\label{fig:geometry}
\end{figure}

Next, we sought to test the suitability of the SCB dynamics to model stem cell dynamics of mammary gland morphogenesis, where extensive cell movements have been reported within tips via intravital live-imaging \cite{Scheele:2017}, with rapid rearrangements occurring on time scales of a few hours (Fig. \ref{fig:geometry} and Fig. S6A of the appendix). In this case, however, tips cannot be followed for long-enough for survival probabilities to be directed measured as in Figs. \ref{fig:RealSurvival} and \ref{fig:RealSurvival_Kidney} for intestine and kidney, respectively. However, extensive clonal dispersion has been observed in quantitative clonal lineage tracing experiments during pubertal growth \cite{Davis:2016, Scheele:2017}, and we therefore sought to infer the value of noise from these experiments (Fig. S6B o the appendix)

Turning back to published lineage-tracing datasets, where single mammary stem cells are labelled at the beginning of puberty (3 weeks of age) and traced until either 5 weeks or 8 weeks of age, clones in tips displayed extensive fragmentation, which is expected to be directly related to the ratio $k_r/k_d$ (Fig. \ref{fig:geometry}c,d and Fig. S6B-D of the appendix). We thus ran as above 2-dimensional simulations of our SCB dynamics (see section 5 of the appendix for details), using measured values of the tip width and length to set the geometry. As a metric for clonal dispersion, we then computationally measured for each labelled cell the distance to its closest clonal neighbour: for a fully cohesive clone, all cells should be touching and the distance to the closest neighbour should be always one cell diameter. Increasing the value of $k_r/k_d$ robustly increased the closest neighbour distance. We then performed the same measurements in the experimental data set, both for the 5 weeks and 8 weeks time points (Fig. \ref{fig:geometry}c,d), and also for luminal and basal cell types separately, given the dominant unipotency of these cell populations in pubertal development \cite{Davis:2016, Scheele:2017, lilja2018clonal, wuidart2018early}. We found highly consistent results in all four cases (average closest distance of around $1.85$ cell diameter) which allowed us to infer a ratio of (see section 5 and figure S6 of the appendix for details):
\begin{equation}
k_r/k_d \approx 2-5\quad,
\label{eq:mammary}
\end{equation}
in mammary gland, emphasizing the importance of considering stochasticity in the conveyor belt picture. Indeed, we found that, with this fitting parameter, the model reproduced well the probability distribution of closest distances, both at the 5 weeks and 8 weeks time points (Fig. \ref{fig:geometry}c,d).

In addition to this value, we must again pay attention to the geometry of the mammary tip, with basal cells forming a 2-dimensional monolayer (similar to the previous cases) while luminal cells form multiple layers in 3 dimensions within the tip. Assuming that the intercalation between cells occurs mainly at the same layer, the system of luminal cells in the tip of the mammary gland can be abstracted as $R-1$ successive hemispherical 2-dimensional layers. Let us emphasize the dependence of $N^{2D}$,  as defined by equation (\ref{eq:Ns}), on $R$, writing $N^{2D}_s\equiv N^{2D}_s(R)$. In that case, the amount of luminal stem cells can be inferred as:
\begin{equation}
N_s^{3D}=\sum_{R'<R}N_s^{2D}(R')\quad.
\label{eq:Ns3D}
\end{equation}
Taking the fitted range of $k_r/k_d\in (2,5)$, together with an estimation of the radius of $R=5\pm 2$, Eq. \ref{eq:Ns3D} then predicts that a number of luminal stem cells per tip of $N_s^{3D}=170\pm 110$, in good quantitative agreement with experimental estimates from lineage tracing of $N_s^{\rm exp}=172 \pm 102$ (mean$\pm$s.d.) \cite{Scheele:2017}. For basal cells, using the same parameters for a 2-dimensional monolayer, Eq. \ref{eq:Ns} predicts that $N_s^{2D}=37\pm 11$, against empirical observations reporting an amount of basal stem cells of at least 15 \cite{Davis:2016}, and $N_s^{\rm exp}=93 \pm 76$  (mean$\pm$s.d.) \cite{Scheele:2017}. Although the prediction thus falls in the correct range, the under-estimation of basal stem cell number may be due to the highly anisotropic geometry of basal stem cells.

 \section*{Discussion}

The main objective of this study was to provide new insights to the question of whether stem cell function is a cell-intrinsic, inherited, property, or rather an extrinsic, context-dependent notion emerging from the collective dynamics of a tissue \cite{Simons:2011, Blanpain:2013, Moris:2016, Yang:2017}. To that end, we took a complementary standpoint to the one based on the classification of molecular markers and their potential functional role, adopting a purely dynamical/geometrical approach - although combining the two would be a logical extension for future work. We analyzed stem cell lineage survival as a purely dynamical process of competition for finite niche space, taking into account the presence of stochastic cell rearrangements, cell duplication rates and tissue geometry. The combination of these ingredients gives rise to a complex reaction-diffusion process that can be abstracted as a stochastic-conveyor belt. We show that survival probability as a function of starting position away from the most favourable position adopts a simple universal Gaussian shape, so that a well-defined number of functional stem cells (i.e. cells which have a non-negligible probability of surviving long-term) arises in the theory, set by tissue geometry and the ratio between random reallocation and cell proliferation rates, $k_r/k_d$. 
We applied this theory to recent live-imaging data tracing stem cell survival as a function of position in the homeostatic intestinal crypt and kidney morphogenesis, and find good quantitative agreement. We also use the model to infer values of $k_r/k_d$ from fixed lineage tracing experiments in mammary gland morphogenesis, and show that this inference allows us to predict the typical number of stem cells in this system. Interestingly, the ratio of noise to advection $k_r/k_d$ appeared to be an order of magnitude larger in kidney development as compared to the intestinal crypt (with mammary gland being intermediate), which explained well the widely different number of functional stem cells observed in each.

Although we have sketched here the simplest source of noise in cellular movements (random exchange of position in cell neighbors), our approach and results are in fact highly robust to different types of microscopic assumptions, and should thus be seen as representative of a general class of models for stem cell dynamics with  advection and noise, rather than a specific microscopic mechanism. In mammary gland and kidney morphogenesis, direct cell-cell rearrangements are observed \cite{Scheele:2017, Riccio:2016}, while kidney also displays mitotic dispersal \cite{Packard:2013}, where noise arises from the randomness of cell re-insertion in the layer after division. 
Furthermore, on short-time scales, directed cellular movements have been observed in kidney tip morphogenesis, with Ret and Etv4 mutant clones being statistically overtaken by wild-type cells, leading to the proposal that Ret/Etv4 were involved in directional movement towards tips \cite{Riccio:2016}. However, tips maintain heterogeneity in Ret expression through branching, arguing that cells must shuttle between high-Ret and low-Ret states \cite{Riccio:2016}, which would effectively contribute to movement stochasticity on long time scales. 
Finally, "tectonic" movements, which collectively reposition cells towards/away from niches, can also be captured (Fig. S5 of the appendix). These are particularly relevant in developmental settings, such as gut morphogenesis, where the global shape of the epithelium changes, displacing collectively cells from villus to crypt regions \cite{Guiu:2019}, or upon tip-spitting during branching morphogenesis \cite{Scheele:2017}. Active migration, as observed in adult intestinal homeostasis \cite{Krndija:2019} could also contribute to such collective random repositioning events. In the future, it would be interesting to further understand quantitatively random cell re-arrangements $k_r$, and how they could be modulated by parameters such as tissue density, aspect ratio, active cell migration or division rates (see section 3C of the appendix for details). Mechanical models of cell motility upon rheological transitions \cite{Bi:2014, Garcia:2015, Popovic:2020}, or of re-arrangements and junctional remodelling upon cell divisions \cite{firmino2016cell, pinheiro2017transmission} in densely packed tissues could also help to understand quantitatively what sets $k_r$ in each system.

The proposed framework can, in principle,  be applied to any tissue dynamics in which niche signals and/or cellular proliferation is localized, leading to directional flows \cite{Moad:2017, Han:2019}. Cases where substantial changes to the model would be necessary are in the context of an "open niche" such as spermatogenesis \cite{Jorg:2019} or skin homeostasis \cite{Rompolas:2016}, where renewing cells form a 2-dimensional layer of neutrally competing equipotent progenitors, thus with little in-plane cellular flows. Finally, the theory could be extended to cases of non-neutral growth. Live-imaging of skin tumor growth for instance is consistent with very low values of $k_r/k_d$ \cite{Brown:2017}, as little to none clonal dispersion is observed, which would tend to favor deterministic growth in our model. Nevertheless, this does not occur as tumor cells trigger higher proliferation rates of normal cells \cite{Brown:2017}, resulting in complex geometrical changes and encapsulation of the malignant clone. Incorporating these types of feedbacks between multiple cell populations \cite{Zhou:2018} in our model would thus have particular relevance to understand the dynamics of tumor initiation\cite{Sanchez:2016}. Our approach must be taken as part of a more general enterprise, namely, the role of both intrinsic cues, and complex collective dynamics in defining the functional stem cells.

\section*{Materials and Methods}

Additional information on the theoretical, computational and experimental methods used can be found in the appendix.

\section*{Acknowledgements}
We thank all members of the Hannezo, Simons and Van Rheenen groups for stimulating discussions.  This work was financially supported by the European Research Council Grant CANCERRECURRENCE 648804 (to J.v.R.), the CancerGenomics.nl (Netherlands Organisation for Scientific Research) program (to J.v.R.), the Doctor Josef Steiner Foundation (to J.v.R). B.D.S acknowledges funding
from the Royal Society E.P. Abraham Research Professorship (RP/R1/180165) and Wellcome Trust (098357/Z/12/Z). 

\noindent
\rule{0.485\textwidth}{0.4pt}



\renewcommand{\thefigure}{S\arabic{figure}}
\renewcommand{\theequation}{S\arabic{equation}}
\setcounter{equation}{0}


\section*{Appendix}
\section{Basic dynamics}

The simplest abstraction of our system is composed by a 1-dimensional column of cells arranged in a finite segment $[0,N]$, in which the length unit is scaled to the average length of a cell. In addition, in this 1-dimensional setting, it is necessary to additionally assume that at $0$ there is a rigid wall. Each cell divides at constant rate $k_d$. Due to the rigid boundary at the bottom, cell proliferation generates a net flow of cells towards the positive axis. In addition, the position of the cells can fluctuate stochastically at rate $k_r$ (either via local cell-cell rearrangements, or more global movements of cells relative to the niche, see sections below for more details). At $t=0$, each cell defines a lineage, labeled by the position they occupy, i.e., the cell located at position $n$ at $t=0$ will be the {\em mother} of the lineage $c_n$. From here on, due to proliferation, drift and random reallocations, we can potentially find cells of lineage $c_n$ in all regions of the organ.  However, as soon as a cell reaches the position $N$ and is pushed, it disappears from the system, eventually resulting in the disappearance of a whole lineage from the system. Notice that, if more than one lineage is present, for a sufficiently long time, the probability of lineage loss is always greater than zero. Since the lineage loss is an irreversible event, that means that one expects that the system will sooner or later become monoclonal, meaning that a single lineage populates the whole system. This competitive dynamics can be metaphorically understood as a conveyor belt with random fluctuations in the cell positions, sketched in Fig. 1 of the main text. This why we call it {\em Stochastic Conveyor Belt} (SCB) dynamics.

\begin{figure}
\begin{center}
\includegraphics[width= 8.5cm]{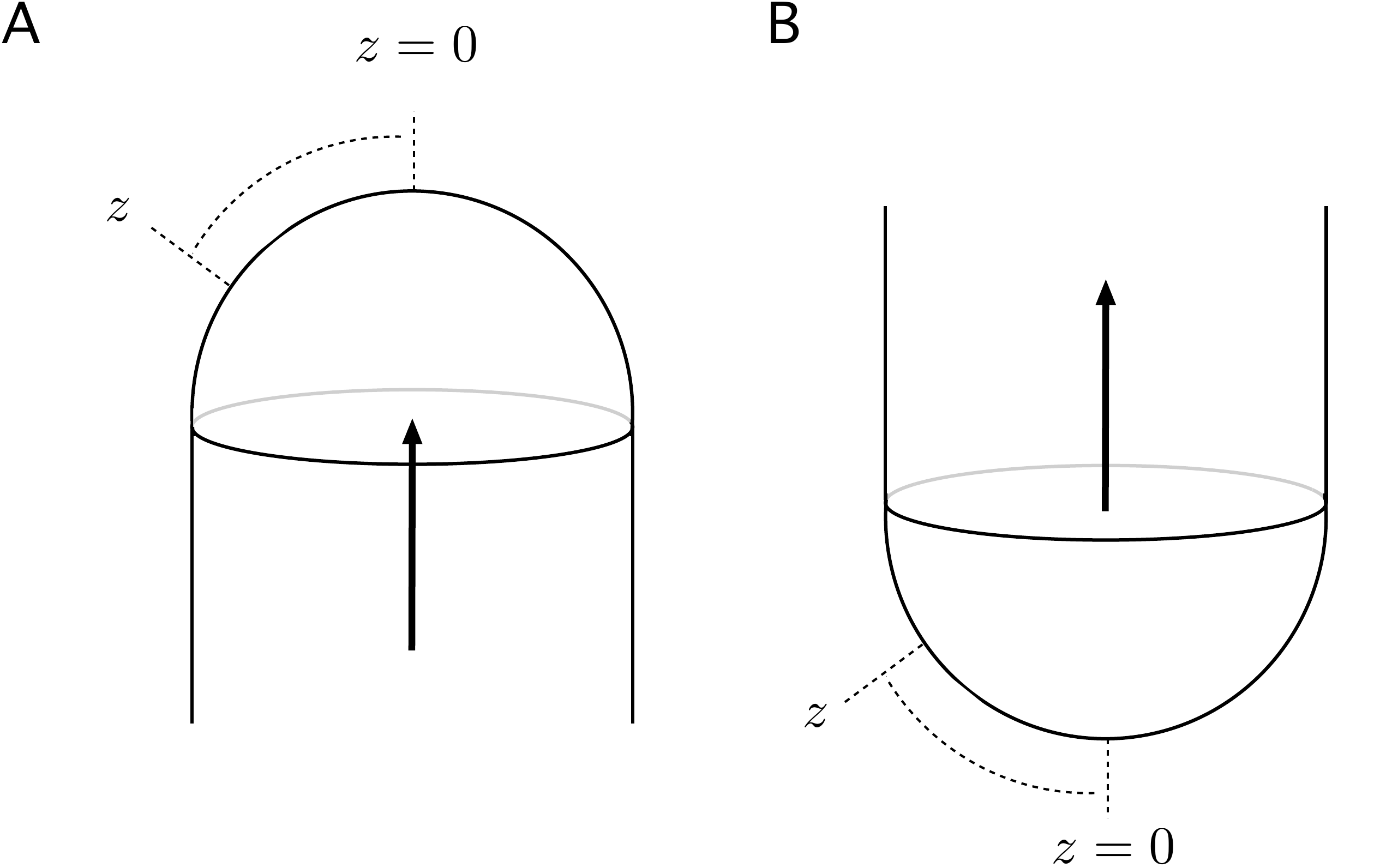}
\caption{The SCB dynamics can describe several process of development and renewal, up to a change on referential frame. A/ In the case of both the mammary gland and kidney development, the organ generates new cells via tip growth while ductal cells remain stationnary, leading to relative tip movement. B/ In the case of the self-renewal of the crypt, the bottom of the system is stationary and cells are flowing upwards}
\label{fig:Frame}
\end{center}
\end{figure}

Both intestinal crypts and elongating mammary or kidney tips can be described by the SCB dynamics, if we consider that it develops over more realistic geometries, such as hemispheric surfaces. An hemispheric surface has the coordinate origin at the bottom/top pole (the tip in the kidney and mammary gland buds, and the bottom of the crypt), and there is no need to assume rigid boundaries anywhere, as the symmetry of the geometry will trigger the SCB dynamics, producing a net flow of cells towards the positive axis $z$, which defines the direction the organ grows/renews. Qualitatively, the dynamics can be abstracted in the same way in the three organs, up to a change of referential frame: intestinal crypts expel differentiated cells with the crypt base staying stationary, while differentiated cells stay in place while the tip moves forward in the case of mammary gland or kidney development --see figure (\ref{fig:Frame}). Importantly, the results that we derive, as described below, are generic to different types of assumptions on the microscopic dynamics, as long as the basis features of advection and rearrangements exist in a system (see sections below for details and examples). 

\subsection{Lineage prevalence evolution}

A full model of the lineage dynamics would require keeping track of each lineage as cell divide and re-arrange (given the constraint that only one cell can occupy a given position of the tissue), leading to an extremely intertwined dynamical system where a few or no analytical predictions can be derived. To overcome this problem, we first make a mean-field approximation, whose suitability will be validated in subsequent section from numerical simulations. We consider a continuous array of cells of length $N$ in which each point (cell) is subject to the SCB dynamics. Given a position $x$, the effect of proliferation at lower levels is captured by an average (mean-field) push-up term, $\sim k_d x$, enabling us to decouple the evolution of different lineages and derive analytical predictions for the prevalence of a given lineage (i.e. cells arising from an ancestor which started at position $n$) in the tissue. We characterize the presence of cells of a given lineage $c_n$ at a given point of the interval $[0,N]$ by a prevalence function $\rho_n(x,t)$ (analogous to an average density). To construct the equation accounting for the time evolution of the prevalence of lineage $c_n$, we need to consider:
\begin{itemize}
\item
{\em Drift term} accounting for the mean field push up movement at position $x$ due to proliferation at rate $k_d$ at lower levels:
\[
-k_d\frac{\partial }{\partial x}\left(x\rho_n(x,t)\right)\quad,
\]
since, at positions $[0,. . .,x)$, $k_d x$ new cells will be produced in a time step pushing the cells at position $x$ an average of $k_d x$ positions upwards.
\item
{\em Diffusive term} accounting for the random reallocations of cells occurring at rate $k_r$:
\[
+\frac{k_r}{2}\frac{\partial^2}{\partial x^2}\rho_n(x,t)\quad.
\]
\item
{\em Proliferative term} accounting for the exponential proliferation of the cells of the lineage under study at rate $k_d$:
\[
+k_d\rho_n(x,t)\quad.
\]
\end{itemize}

Combining these terms, the equation accounting for the time evolution (in units of $\tau=k_d t$) of such prevalence is:
\begin{equation}
\frac{\partial \rho_n}{\partial \tau}=-\frac{\partial }{\partial x}(x\rho_n)+\frac{k_r}{2k_d}\frac{\partial^2 \rho_n}{\partial x^2}+\rho_n\quad.
\label{eq:FisherKPP}
\end{equation}
We will refer to this equation as the {\em SCB equation}.

\subsection{Solutions of the SCB equation}

To solve equation (\ref{eq:FisherKPP}) we impose as initial condition that $\rho_n(x,0)\sim\mathbf{N}(n,1/2)$, i.e., a normal distribution centred $n$ with variance $1/2$. 
This condition describes a single cell located at position $n$ as a density that spreads significantly only at $n\pm 1/2$ and whose integral is equal to $1$, as it is expected for a single cell. Natural boundary conditions apply\footnote{A rigorous approach to this problem would require a reflecting boundary condition at $x=0$. Imposing such boundary condition would make the whole problem much more difficult and, eventually intractable. The reason by which we adopted natural boundary conditions is due to the fact that the dynamics in this system is extremely imbalanced and runs essentially in only one direction. If one takes equation (\ref{eq:solutionP}) at $x=0$ we observe that the probability of being at $x=0$ decays as $\sim e^{-\tau}$, for any starting point $n>0$, as it is the case in our system. This tells us that the probability of visiting regions $x<0$ is, to our purposes, negligible.}.
We observe that the solution  of equation (\ref{eq:FisherKPP}) factorizes as follows:
\begin{equation}
\rho_n(x,t)\propto \phi_1(x,t)\phi_2(t)\quad,
\label{eq:Factor}
\end{equation}
where $\phi_1$ is the solution of:
\begin{equation}
\frac{\partial \phi_1}{\partial \tau}=-\frac{\partial }{\partial x}(x\phi_1)+\frac{k_r}{2k_d}\frac{\partial^2\phi_1}{\partial x^2}\quad,
\label{eq:FPPhi_1}
\end{equation}
and $\phi_2$ is the solution of:
\[
\frac{\partial \phi_2}{\partial \tau}=\phi_2\quad.
\]
which has a simple solution:
\begin{equation}
\phi_2\propto e^{\tau}\quad.
\label{eq:exp}
\end{equation}
In words, $\phi_2$ represents the fact that cells proliferate randomly at rate $k_d$ so that the entire lineage increases in size exponentially regardless of its position.

Computing $\phi_1$ is a little bit more entangled. To start with, first observe that (\ref{eq:FPPhi_1}) is a Fokker-Planck like equation for the time evolution of the probabilities of a random variable following a mixture of Brownian motion with a given amplitude $k_r/k_d$ and a drift parameter $x\phi_1(x,\tau)$.  If, instead of looking at the evolution of the probabilities we look at the behaviour of the random variable itself --the position in the system, $X$--, equation (\ref{eq:FPPhi_1}) has its stochastic differential equation counterpart in:
\begin{equation}
dX=Xd\tau+\sqrt{\frac{k_r}{k_d}}dW\quad,
\label{eq:LangevinA}
\end{equation}
being $dW$ the differential of the standard Brownian motion \cite{Gardiner:1984, VanKampen:2007}. Equation (\ref{eq:LangevinA}) describes the movement of a single cell in the SCB dynamics. Eq (\ref{eq:LangevinA}) explicitly tracks the real trajectory of a cell (and not lineage prevalence).
The above described stochastic process has no stationary solutions, which is in agreement with the SCB dynamics: all cells will sooner or later be pushed out from the system (in the case of the crypt, or will be left behind, as in the case of the mammary gland development). Imposing the following initial conditions $t_0=0$, $\phi_1(x,0)\sim\mathbf{N}(n,1/2)$ and natural boundary conditions, we observe that:
\[
d\left(e^{-\tau} X(\tau)\right)=dX(\tau)e^{-\tau}-e^{-\tau}X(\tau)d\tau\quad.
\]
Then, multiplying both sides of equation (\ref{eq:LangevinA}) by $e^{-\tau}$, and after some algebra, one finds that:
\[
d\left(e^{-\tau} X(\tau)\right)=\sqrt{\frac{k_r}{k_d}}e^{-\tau}dW\quad,
\]
leading to:
\begin{equation}
X(\tau)=x_0e^{\tau}+\int_0^\tau \sqrt{\frac{k_r}{k_d}}e^{(\tau-s)} dW\quad.
\label{eq:Solution}
\end{equation}
The integral is a standard stochastic integral with respect to a Wiener process. According to {\em Ito's isometry} \cite{Karatzas:1988} 
one has that the law governing the random variable described by the integral is a normal distribution $\mathbf{N}(0,\sigma^2(\tau))$. In our case this reads: 
\[
\int_0^\tau e^{(\tau-s)}\sqrt{\frac{k_r}{k_d}} dW \sim{\mathbf N}\left(0,\int_0^t\left|\sqrt{\frac{k_r}{k_d}}e^{(t-s)}\right|^2ds\right)\quad,
\]
which means that the explicit form of $\sigma^2(\tau)$, is thus given by:
\begin{eqnarray}
\sigma^2(\tau)=\int_0^\tau\left|\sqrt{\frac{k_r}{k_d}}e^{(\tau-s)}\right|^2ds=\frac{k_r}{2k_d}\left(e^{2\tau}-1\right)\quad. 
\label{eq:sigmaT}
\end{eqnarray}
Finally, from equation (\ref{eq:Solution}), we conclude that the time dependent mean, $\mu(\tau)$, is:
\[
\mu(t, x_0)=x_0e^{\tau}\quad.
\]
That leads to:
\begin{equation}
\phi_1(x,\tau)\propto\sqrt{\frac{k_d}{2\pi k_r\left(e^{2\tau}-1\right)}}{\rm exp}\left\{{-\frac{k_d}{2k_r}\frac{\left(x-ne^{\tau}\right)^2}{\left(e^{2\tau}-1\right)}}\right\}\;.
\label{eq:solutionP}
\end{equation}
In other words, the solution is given by a random variable following a normal distribution whose mean and variance run exponentially fast in time through the positive axis. In Fig. (\ref{fig:Dynamics}a) of this appendix, we plotted some snapshots of this time dependent probability.

According to equations (\ref{eq:Factor}, \ref{eq:exp}) and (\ref{eq:solutionP}), the solution of the SCB equation (\ref{eq:FisherKPP}) can be fairly approximated as:
\begin{equation}
\rho_n(x,\tau)\propto \sqrt{\frac{k_d}{2\pi k_r}}{\rm exp}\left\{-\frac{k_d}{2k_r}\left(\frac{x-ne^{\tau}}{e^{\tau}}\right)^2\right\}\quad.
\label{eq:Safeapprox}
\end{equation}

\subsection{Lineage survival probability}
\label{sec:LSP}

Experimentally, a measurable quantity of key interest is the long-term fixation probability ("how likely is it for a cell starting at a given position $n$ to take over the entire crypt?"). Indeed, at a discrete cellular level, a given lineage can disappear from the conveyor belt (absorbing boundary condition at the end of the belt), so that a single lineage will be present in the crypt on the long term, something that by definition cannot be captured in the continuum model.

\begin{figure}
\begin{center}
\includegraphics[width= 8.5cm]{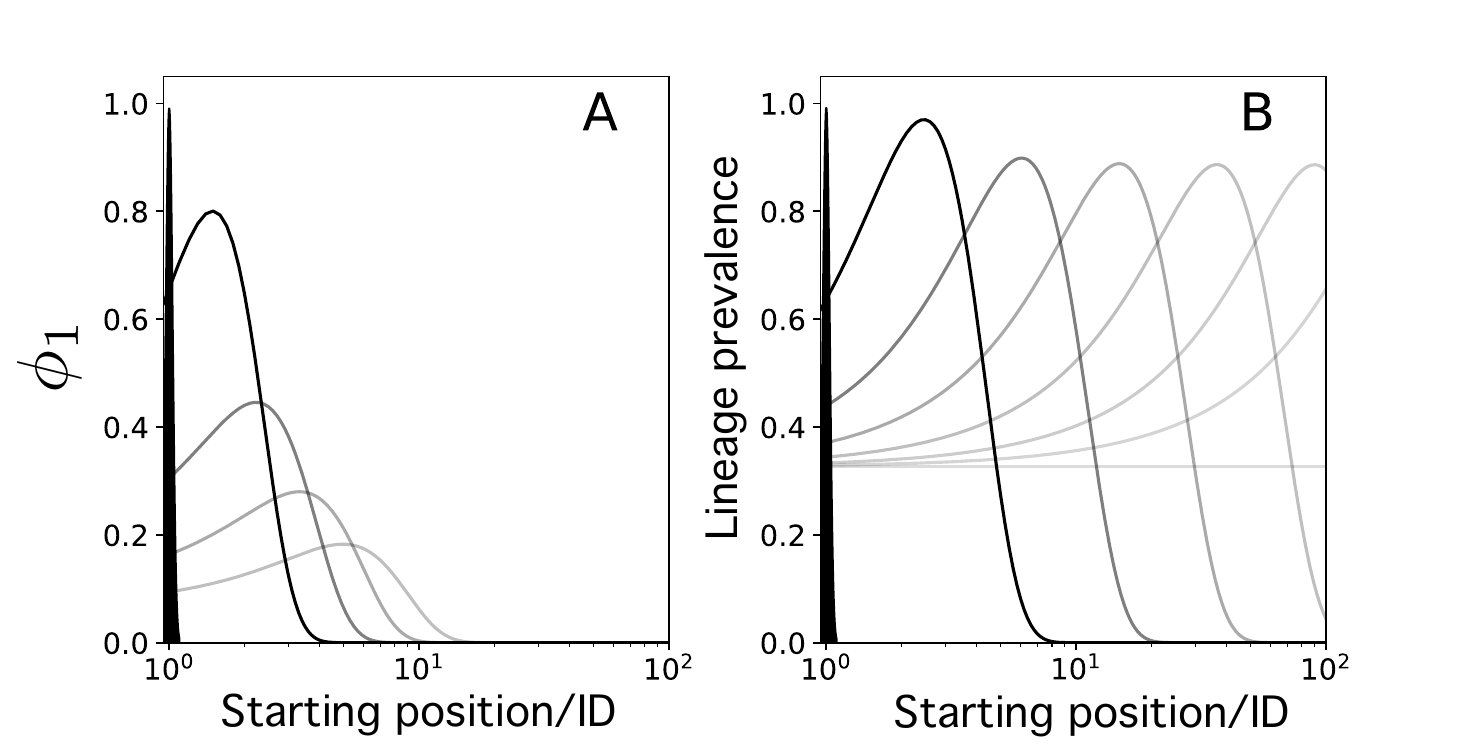}
\caption{Temporal evolution of the SCB dynamics (black to grey indicates time). A/ Evolution in time of the probability for cells starting at a given position $n$ (in that case, $n=1$) to occupy the location $z$ at a given time according to the theoretical prediction given by equation (\ref{eq:solutionP}). Observe that the dynamics does not run to a stationary state, so all cells will eventually abandon the system with probability $1$ as long as time grows. These probability densities refer to real probabilities that the cell is in a given position during the stochastic trajectory defined to the SCB dynamics, in contrast to the relative interpretation of the lineage prevalence. B/ Evolution in time of the prevalence of a lineage starting in the same position across all positions, according to the solution of the SCB equation (\ref{eq:FisherKPP}) given in equation (\ref{eq:Safeapprox}). Observe that the reaction diffusion dynamics displays a front that runs exponentially towards the outside of the system. Notice the prevalence reaches a non-zero stationary value, that is assumed to be proportional to the probability of the lineage to remain and colonize the whole system. The initial black, elongated triangle at position $1$ shows the initial conditions i.e., $n=1$, and $k_r=1, k_d=1$. These values have been chosen only for the sake of clarity.}
\label{fig:Dynamics}
\end{center}
\end{figure}

However, we note that the lineage prevalence from the continuum model converges on long time scales towards a simple scaling law
\begin{eqnarray}
\rho_n(\infty)&\equiv&\lim_{\tau\to \infty}\rho_n(x,\tau)\nonumber\\
&=&\sqrt{\frac{k_d}{2\pi k_r}}e^{-\frac{k_d}{2k_r}n^2}\quad,
\label{eq:rhoinf=}
\end{eqnarray}
which is independent of $x$ and $t$ (see Fig. (\ref{fig:Dynamics}b) of this appendix), and dependent only of the ratio kr/kd and the starting position k. This argues that on the long-term, lineages starting at different positions $n$ and $n'$ have well-defined relative prevalence. Together with the observation that conveyor belts tend to monoclonality, it is then reasonable to make the assumption that

\begin{itemize}
\item
The long term lineage fixation/survival probability is proportional to the asymptotic lineage prevalence $\rho_n(\infty)$
\end{itemize}

\noindent
so that the probability of lineage survival, $p(c_n)$, can be derived directly from the normalization of the asymptotic prevalences $\rho_n(\infty)$ as:
\begin{equation}
p(c_n)\approx\frac{\rho_n(\infty)}{\sum_j\rho_j(\infty)}\quad,\nonumber
\end{equation}
which leads to:
\begin{equation}
p(c_n)\propto  {\rm exp}\left\{-\frac{k_d}{2k_r}n^2\right\}\quad.
\label{eq:p(c_n)=}
\end{equation}

In words, we predict that the lineage fixation/survival probability is described by a Gaussian-like distribution defined only over the positive axis with mean $0$
and variance $\sqrt{\frac{k_r}{k_d}}$. Importantly, numerical simulations of the full discrete SCB model in 1 dimension revealed excellent agreement with Eq. (\ref{eq:p(c_n)=}), validating our assumptions.

\section{SCB dynamics in more general geometries}
\label{sec:General}
\begin{figure}[ht!]
\begin{center}
\includegraphics[width= 8.5cm]{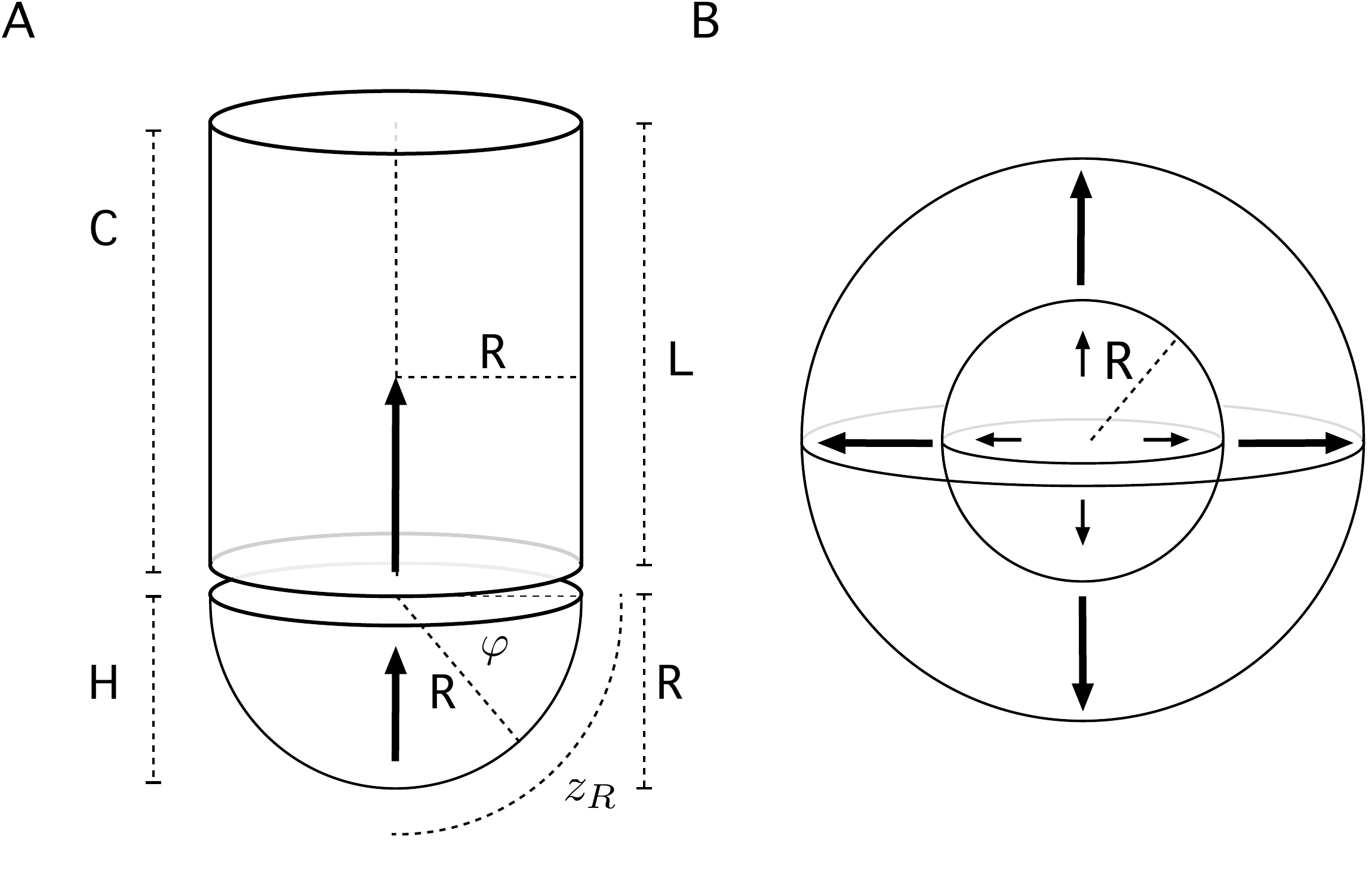}
\caption{A/ Schematic characterization of the structure of the crypt as a hemispherical region $H$ coupled to a cylinder region of the same radius, $R$. B/ The expansion of the tissue in a 3-dimensional abstract setting where there is radial symmetry. The growing of the inner cells creates a push up force. In addition, the stochastic fluctuations in the position determine the probability of lineage survival as a function of the starting point, as in the case of low dimensional approaches. }
\label{fig:CryptGeom}
\end{center}
\end{figure}

In general we will assume that there is a coordinate $z$ over which the displacement induced by proliferation takes place (in practice this would be dictated by the boundary conditions, e.g. the geometry of the region where cell loss occurs). All the dynamics will be, in consequence, studied from its projection over this coordinate --see Fig.  (\ref{fig:CryptGeom}) of this appendixfor the special cases of hemispheric and spheric geometries. 
In the case of a $1$-dimensional system, as the one described above, this coordinate is the length, $x$. In the case of a hemisphere, assuming that the push-up force is exerted from the bottom pole, this coordinate is the arc length defined from the position of the cell to the bottom pole itself --see Fig. (\ref{fig:CryptGeom}a) of this appendix. To gain intuition, consider the surface of the hemisphere with radius $R$: The cells at the bottom pole divide and push the ones on top of them up through the surface. The cell under consideration is located at a position defining an arc from the bottom pole equal to $z=R\varphi_n=z_n$, where $\varphi_n$ is the polar angle, meaning that there is an arc of $n$ cells from the given cell to the pole of the hemisphere --see Fig. (\ref{fig:CryptGeom}a) of this appendix. The successive divisions of cells located at $z_i<z_n$ will result into a net displacement along the polar coordinate $\varphi$ of the cell located initially at $z_n=R\varphi_n$, going from $z_n=R\varphi_n$ to $z_{k'}=R\varphi_n'$, with $\varphi_n'>\varphi_n$. The linear displacement along the surface will be $\Delta z=R(\varphi_n'-\varphi_n)$. Displacements along the other coordinate will have no net effect in the push up force.

\subsection{SCB equation in general geometries} 

\begin{figure*}
\begin{center}
\includegraphics[width= 16cm]{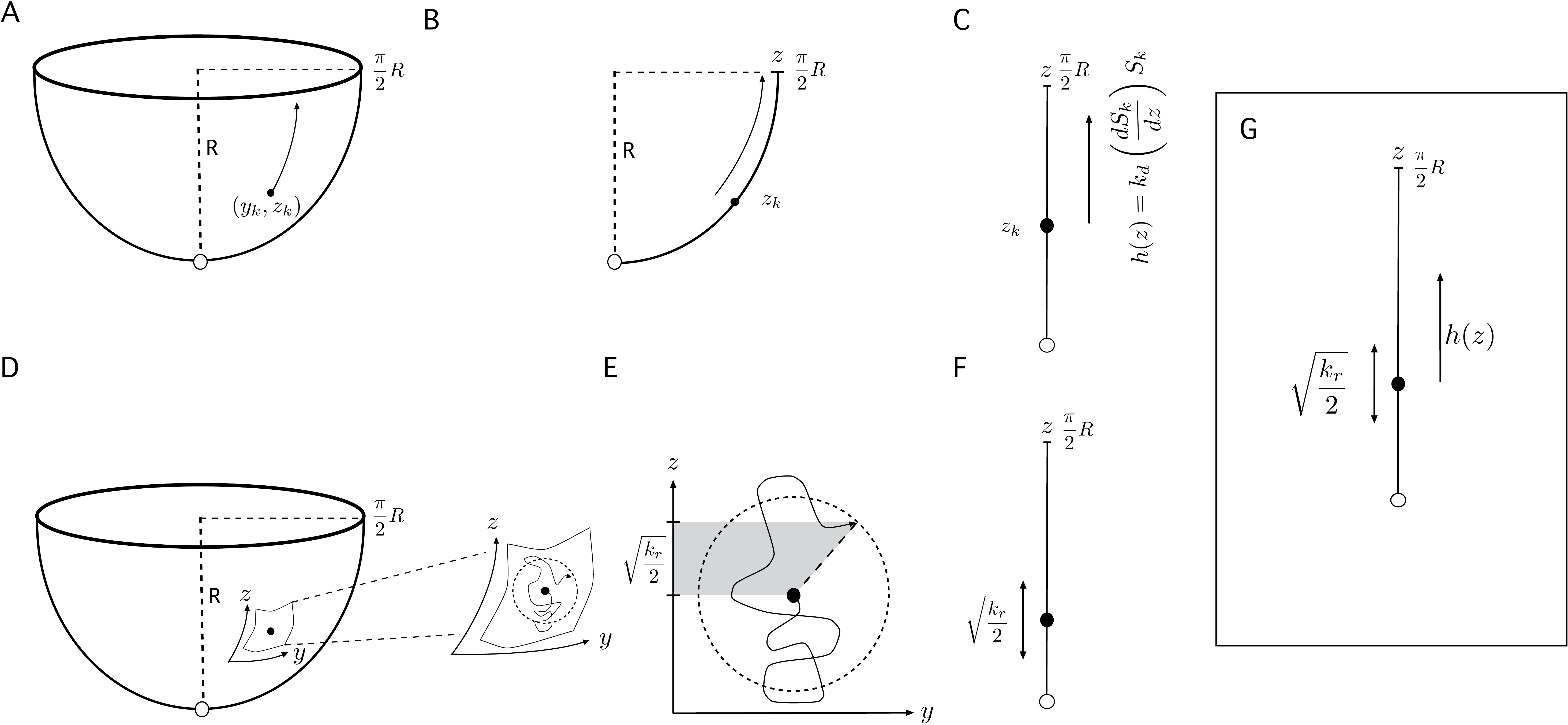}
\caption{
Constructing the dynamical equation for a general kind of manifold and projecting it onto the 1-dimensional coordinate system along which the push-up force is exerted.  A/ A cell is located in a point in a manifold --in that case, a hemisphere--, described by two orthogonal coordinates $(y_n,z_n)$. In this case $y_n=R\theta_n$, where $\theta$ is the azimuthal angle, and $z_n=R\varphi_n$, where $\varphi$ is the polar angle. B/ the push up force due to proliferation is exerted only along the $z$ coordinate. C/ $h(z)$ is the drift term that enters the equation, and refers to the amount of new surface that has been created below the point $z_n$ in the $z$ coordinate that results in pushing up the cells above. D/ the same point $(y_n,z_n)$ also observes fluctuations due to random noise. In particular, the rate of these fluctuations is externally reported as $k_r$. E/ Thanks to the local flatness property, if $\sqrt{k_r}\ll R$, then the fluctuations take place locally in a flat space, and the average distance from the starting point, after a unit time interval, will be $\sqrt{k_r}$. Since the coordinate system $y,z$ is orthogonal and locally flat, and the random fluctuations occur isotropically in space, the projection of the fluctuations over the coordinate $y$, $k^y_r$, will be the same than in the coordinate $z$, $k_r^z$. Since $\sqrt{k_r}=\sqrt{k_r^y+k_r^z}$, the only solution for this projection is that, $k_r^z=k_r/2$, as described in F/. G/ Combining C/ and F/ we have that the growing process can be described as a SC dynamics along the $z$ coordinate with drift term $h(z)$, diffusion term $\frac{k_r}{2}$, and proliferative term $\rho_n z$.}
\label{fig:Fluct}
\end{center}
\end{figure*}

\subsubsection{Drift term}
The drift term will be described by the function $h(z)$, accounting for the local speed of a cell at position $z$:
\begin{equation}
h(z)\equiv\frac{dz}{dt}\quad.
\label{eq:Defh}
\end{equation}
In the case of a $1$-dimensional system, as the one described by equation (\ref{eq:FisherKPP}), one has that $h(x)=k_d x$. 

To properly study this dynamics over more general geometries, let us consider a Riemannian manifold equipped with a metric tensor $\mathbf{g}$, with components $g_{ij}$ \cite{Klingenberg:1978}. Crucial to our purposes is the property of {\em local flatness} \cite{Klingenberg:1978, Kuhnel:2006}. Roughly speaking, this implies that, for small enough regions of the manifold, the geometry has euclidean properties. Let us consider that the push up force due to proliferation has an origin and is exerted along the direction of a single coordinate $z$ as well. 
As we did above, the surface/volume units are given such that an average cell has a surface/volume of $1$ in the corresponding units. Consider the starting position of our cell to be $z_n$ along the coordinate $z$. If the other coordinates of the manifold are $x_1, . . .,x_{n-1}$, the surface/volume encapsulated {\em below} position $z_n$ is given by:
\[
S_n=\int . . .\int_0^{z_n}\sqrt{g}dx_1. . . dx_{n-1}d{z}\quad,
\]
where $g$ is the determiner of the metric tensor, i.e.:
\[
g=\left|
\begin{array}{lllll}
g_{11} & g_{12} &. & .&. \\
g_{21} & g_{22} & & & \\
 .&  & . &  & \\
 .&  & &. &\\
 .&  & & &.\\
 \end{array}
\right|
\]
Cells are assumed to divide at rate $k_d$. That implies that $k_d S_n$ new cells will be produced {\em below} the cell located at $z_n$. This will create an extra surface/volume of:
\[
\frac{dS_n}{dt}=k_d S_n\quad,
\]
that will project into the coordinate $z$. Using that:
\[
\frac{dS_n}{dt}=\frac{dS_n}{dz}\frac{dz}{dt}\quad, 
\]
and, then, equation (\ref{eq:Defh}), one can find the general expression for this projection, which reads:
\begin{equation}
h(z)=k_d\left(\frac{dS_n}{dz}\right)^{-1}S_n\quad.
\label{eq:h}
\end{equation}

\subsubsection{Diffusion term}

We are only interested on the projection of the dynamics over the coordinate $z$ along which the system grows, as in the other coordinates the  competition is neutral and has no net effect in the lineage survival statistics.  Under the assumption of local flatness, if the global stochastic reallocation rate is isotropic one can estimate the projection of such reallocations over the coordinate $z$, $k_r^z$ as:
\begin{equation}
k_r^z\sim \frac{k_r}{D}\quad,
\label{eq:krz}
\end{equation}
where $D$ is the dimension of the manifold. 

Let us consider in detail the 2-dimensional case: Imagine that we report experimental, isotropic fluctuations of amplitude $k_r$ (see Fig. (\ref{fig:Fluct})) of this appendix. That is, in a time unit, the cells move randomly over the manifold $k_r$ steps. We have a $2$-dimensional isotropic dynamics taking place over a locally flat \cite{Kuhnel:2006} surface with generic orthogonal coordinates $y,z$ (for example, $R\times$ the azimuthal angle $\theta$ and $R\times$ the polar angle $\varphi$ over a sphere surface). The amplitude of the fluctuations after time $t$ is known to be $\sim \sqrt{k_rt}$, a distance defined over the surface. In the case we consider the projection over the coordinate $z$, thanks to the local flatness, assuming that $\sqrt{k_r}\ll R$ and using only symmetry reasonings, one has that since the displacement is given by $(\Delta y,\Delta z)=(\sqrt{k_r^y},\sqrt{k_r^z})$:
\[
\sqrt{(\sqrt{k_r^y})^2+(\sqrt{k_r^z})^2}=\sqrt{k_r}\quad.
\]
Since the fluctuations are isotropic:
\[
k_r^z=k_r^y\quad,
\]
leading to:
\begin{equation}
k^z_r=\frac{k_r}{2}\quad.
\label{eq:ukz}
\end{equation}
In the case we are dealing with a spherical surface, we are projecting the fluctuations over the polar angle $z=R\varphi$ (see Fig. (\ref{fig:Fluct}) of this appendix).

\subsubsection{SCB equation}
According to the above results, we will have that the general equation for the evolution of cell lineage prevalences along the coordinate $z$ will read:
\begin{equation}
\frac{\partial \rho_n}{\partial t}=-\frac{\partial }{\partial z}(h(z)\rho_n)+\frac{k^z_r}{2}\frac{\partial^2 \rho_n}{\partial z^2}+\rho_n\quad.
\label{eq:FisherKPPGen}
\end{equation}
In the case $h(z)$ can be approached as a linear function, i.e.,  $h(z)\sim a k_d z$ and $k^z_r$ as $k^z_r=bk_r$, one can apply the assumption presented in section 1\ref{sec:LSP} and rewrite equation (\ref{eq:p(c_n)=}) as:
\begin{equation}
p(c_n)\propto {\rm exp}\left\{-\frac{a}{2b}\frac{k_d}{k_r}n^2\right\}\quad.
\label{eq:very_general_G}
\end{equation}

\subsection{SCB dynamics in realistic geometries}

Let us now consider a detailed version of the geometry of the organs under study in the main text (mammary and kidney tips, or intestinal crypts). They are described as a half sphere, $H$, whose arc length from the bottom pole to the end is is $z _R=\frac{\pi}{2}R$ coupled to a cylinder $C$ of length $L$ and radius $R$. The push-up force is directed towards the top of the cylinder (see Fig. (\ref{fig:CryptGeom}a)) of this appendix. In the arc that goes from the bottom pole to the end of the hemisphere there are $z_R$ cells. Again, the units are given considering the average size of the cell as the 
length/surface/volume unit. Therefore, the cells will be labelled in terms of the geodesic distance over the hemisphere to the bottom pole. 

\subsubsection{Hemispheric region}
The prevalence of the lineages will be given by the $\approx\frac{\pi}{2}R$ cells that populate the arc length that goes from the bottom pole of the hemisphere to the equator, where the system is coupled to the cyclinder. Let us label them as:
\[
(\rho_0,\rho_1,\rho_2, . . .,\rho_{z_{R-1}},\rho_{z_R} )\quad.
\]
Since the coordinate $R$ is constant, the only dynamically relevant information will come from the angle $\varphi$. Each position $n$ in the arc $(0,1,2, . . .,z_n,. . .,z_R-1,z_R)$ describing the initial point of a cell lineage can be rewritten as:
\[
z_n=R\varphi_n\;,\quad \varphi_n\in\left(0,\frac{\pi}{2}\right)\quad.
\]
i.e., $\varphi_n=\frac{z_n}{R}$.  The metric tensor for this hemispheric surface is  \cite{Kuhnel:2006}:
\[
{\mathbf g}=\left(
\begin{array}{cc}
R^2 & 0\\
0 &R^2\sin^2(\varphi)
\end{array}
\right)\quad.
\]
Computing the determiner of ${\mathbf g}$, $g$:
\[
 g=\left|
\begin{array}{cc}
R^2 & 0\\
0 &R^2\sin^2(\varphi)
\end{array}
\right|=R^4\sin^2\varphi\quad,
\]
one can compute the surface element as \cite{Kuhnel:2006}:
\[
dS=\sqrt{g}d\theta d\varphi\quad.
\]
In consequence, the area under the position of the cell $n$ in the  in the hemisphere $H$, located at the arc position $z_n$, will be:
\[
S^H_n=\int_0^{2\pi}\int_0^{\frac{z_n}{R}}\sqrt{g}d\theta d\varphi=2\pi R^2\left(1-\cos\left( \frac{z_n}{R}\right)\right)\quad.
\]
By direct application of equation (\ref{eq:h}), we have that the push-up force inside the hemisphere $H$ is given by:
\begin{equation}
h^H(z)=k_d R\left[\frac{1-\cos\left(\frac{z_n}{R}\right)}{\sin \left(\frac{z_n}{R}\right)}\right]\quad.
\label{eq:h^H}
\end{equation}
Finally, from equation (\ref{eq:ukz}) we know that (see also Fig. (\ref{fig:Fluct}) of this appendix):
\[
k_r^z=\frac{k_r}{2}\quad,
\]
leading, according to the SCB equation (\ref{eq:FisherKPPGen}) for $0\leq z\leq \frac{\pi}{2}R$ in a hemispherical surface to be:
\begin{equation}
\frac{\partial \rho_n}{\partial t}=-k_d R\frac{\partial }{\partial z}\left(  \frac{1-\cos\left(\frac{z_n}{R}\right)}{\sin \left(\frac{z_n}{R}\right)}\rho_n\right)+\frac{k_r}{4}\frac{\partial^2 \rho_n}{\partial z^2}+\rho_n\quad.
\label{eq:FisherKPPSphere}
\end{equation}
The above equation is difficult to deal with. However, we observe that in the region of interest, $z\in\left[0,\frac{\pi}{2}R\right]$, equation (\ref{eq:h^H}) can be approximated as:
\begin{equation}
	\tilde{h}(z)\sim \frac{2k_d}{\pi}z\quad,
\end{equation}
leading to an error bounded as:
\[
\max_{z\in\left[0,\frac{\pi}{2} R\right]}\left |\left | h(z)-\frac{2k_d}{\pi}z\right |\right |<0.09 k_dR\quad,
\]
according to numerical tests. With this approximation, we have that SCB equation for hemispheric surfaces (\ref{eq:FisherKPPSphere}) can be rewritten approximately as:
\begin{equation}
\frac{\partial \rho_n}{\partial t}\approx-\frac{2k_d}{\pi} \frac{\partial }{\partial z}(z\rho_n)+\frac{k_r}{4}\frac{\partial^2 \rho_n}{\partial z^2}+\rho_n\quad,
\label{eq:FisherKPPapprox}
\end{equation}
which is the general kind of SCB equations we have been working so far.

\subsubsection{Coupling to a cylinder}

In the case the cell is at the position $n$ in the cylindric region $C$, the area under it will be given by $S^H_{\frac{\pi}{2}R}$, the area of the whole hemisphere, and the remaining surface due to the cell is the cylinder. Knowing that for the cylindric coordinates $\sqrt{g}=R$, then:
\[
S^C_n=S^H_{\frac{\pi}{2}R}+\int_0^{2\pi} d\theta \int^{z_n}_{\frac{\pi}{2}R}\sqrt{g}dz=S^H_{\frac{\pi}{2}R}+2\pi R\left (z_n-\frac{\pi}{2}R\right)\quad.
\]
Completing the picture, the push force felt by a cell in the cylindric region $C$ is given by:
\begin{equation}
h^C(z)=k_d\left( z_n+\left(1-\frac{\pi}{2}\right)R\right)\quad.
\label{eq:h^C}
\end{equation}
It is easy to check that:
\begin{eqnarray}
\lim_{z\to \frac{\pi}{2}R^+}h^H(z)&=&\lim_{z\to \frac{\pi}{2}R^-}h^C(z)\nonumber\\
\lim_{z\to \frac{\pi}{2}R^+}\frac{d}{dz}h^H(z)&=&\lim_{z\to \frac{\pi}{2}R^-}\frac{d}{dz}h^C(z)\quad.\nonumber
\end{eqnarray}
Therefore, one can define a function, $h(z)$ as:
\begin{eqnarray}
h^{H,C}(z)=\left\{
\begin{array}{ll}
h^H(z)\;{\rm if}\;z\leq \frac{\pi}{2}R\\
h^C(z)\;{\rm if}\;z>\frac{\pi}{2}R\quad,
\end{array}
\right.
\label{eq:h(HC)}
\end{eqnarray}
which is always well defined. In addition, the projection of the fluctuations will be the same in both regions, since the only relevant property is the local dimension, which in both cases is $2$, leading to a $k_r^z=k_r/2$. Consequently, equation (\ref{eq:FisherKPPGen}) can be rewritten consistently for all the hemisphere/cylinder system as:
\begin{equation}
\frac{\partial \rho_n}{\partial t}=-\frac{\partial }{\partial z}(h^{H,C}(z)\rho_n)+\frac{k_r}{4}\frac{\partial^2 \rho_n}{\partial z^2}+\rho_n\quad.
\label{eq:FisherKPPHC}
\end{equation}

\subsubsection{Lineage survival probability in hemispheric geometries}
We then estimate the probability of lineage survival in the limit where most of the cells with non-vanishing long-term survival probability are located in the hemispheric region $H$ of the crypts (putting an upper bound of $k_r/k_d$). As discussed in section \ref{sec:General} of this appendix, the existence of linear functions approximating the drift and fluctuation parameters of the general reaction-diffusion equation (\ref{eq:FisherKPPGen}) leads to Gaussian-like lineage survival probabilities. According to equation (\ref{eq:FisherKPPapprox}) and the assumption on lineage survival probabilities presented in section 1\ref{sec:LSP}, we have that, considering only the hemispheric region of the crypt:
\begin{equation}
p(c_n)\propto {\rm exp}\left\{-\frac{2}{\pi}\frac{k_d}{k_r}n^2\right\}\quad.
\label{eq:pckhemisphere}
\end{equation}

\subsubsection{Higher dimensionality with radial symmetry}
We next consider the case in which the tissue has radial symmetry (i.e. can be abstracted as a manifold isomorphic to a disk or sphere). In consequence, we will only consider the $r$ coordinate away from the center of the tissue (see Fig. (\ref{fig:CryptGeom}b) of this appendix). In the case of isotropic stochastic reallocations occurring at rate  $k_r$, following the same reasoning we used in section \ref{sec:General}, we can approximate the effective rate projected towards the $r$ axis by:
\[
k_r^r\sim\frac{k_r}{{D}}\quad,
\]
where $D$ is the dimension of the manifold. The diffusion term in the reaction diffusion equation will be given by:
\[
\sim \frac{k_r}{{D}}\frac{\partial^2 \rho_n}{\partial r^2}\quad,
\]
Now we compute the drift term. For that, we apply directly equation (\ref{eq:h}). In the case of a $3$-dimensional sphere growing from the centre, we have, if $V(r)$ is the volume encapsulated by the surface of radius $r$, the displacement of a cell at this position  will be determined by:
\[
\frac{d V(r)}{dt}=k_d V(r)\quad,
\]
If $V(r)=\frac{4}{3}\pi r^3$, then, according to the above equation:
\[
\frac{d V(r)}{dt}\propto r^2\frac{dr}{dt}\quad.
\]
 Thus, the push up force will, accordingly, result in a radial displacement of:
\[
\frac{dr}{dt}\propto k_d r\;,\quad h(r)\propto k_d r\quad.
\]
The above speed will define the drift term $h(r)$.

According to the above results, the SCB equation for a spherical system ($D=3$) dividing through the coordinate $r$ will be:
\[
\frac{\partial \rho_n}{\partial t}=-\frac{\partial }{\partial r}(k_d r\rho_n)+\frac{k_r}{6}\frac{\partial^2 \rho_n}{\partial r^2}+\rho_n\quad.
\]
where there must be a boundary region where the cells do not proliferate anymore or a $R_{\rm max}$ beyond which cells are lost, in order to balance this proliferative flux. 

\section{Number of stem cells emerging from the SCB dynamics}
The existence of Gaussian-like distributions in lineage survival probabilities enables us to compute the size of the emerging stem-cell region as the number of cells that are at a distance $\leq \sigma$ from the origin. The reason is that they represent the set of cells having the highest chances to colonize the whole system. In consequence, the origin of stem cell potential in this model is purely dynamical, and the functional stem cell number will arise simply from the interplay between geometry and dynamics. 

\subsection{General case}
Consider that, as we did above, that tissues can be abstracted as a $n$-dimensional $(x_1, . . .,x_{n-1},z)$ Riemannian manifold equipped with a metric tensor $\mathbf{g}$ where the local flatness property holds. Assume there is an origin from which the push up force due to proliferation is exerted along the positive direction of a single coordinate $z$, Eq. (\ref{eq:very_general_G}) applies, and a functional stem cell is defined as any cells within a distance $\sigma$ of the niche center, meaning that, there is a {\em belt} containing
\[
\sim\sigma+1\quad
\]
cells from the origin of coordinate $z$ to the last cell that is inside the stem-cell region, with:
\[
\sigma\sim \sqrt{\frac{b}{a}\frac{k_r}{k_d}}\quad,
\]
Therefore, the number of stem cells emerging from the SCB dynamics in general geometries will be:
\begin{equation}
N_s\approx \int . . .\int_0^{\sigma+1}\sqrt{g}dx_1. . . dx_{n-1}d{z}\quad.
\label{eq:N_s_verygeneral}
\end{equation}

\subsection{Specific cases of experimental interest}

\subsubsection{1-dimensional case}
In this case, we recover the solution from (\ref{eq:p(c_n)=}). According to the definition of the stem cell number given in equation (\ref{eq:N_s_verygeneral}), leading to:
\[
N_s^{1D}\approx 1+\sqrt{\frac{k_r}{k_d}}\quad.
\]
In two or more dimensions, the stem cell number does not come solely from the characteristic length set by fluctuations $k_r/k_d$ along the direction of flow, but also from the geometry of the tissue. 

\subsubsection{Cylindric geometry with radius $R$}

Direct application of equation (\ref{eq:N_s_verygeneral}) on cylindric coordinates considering that the radius is $R$ gives us:
\[
N_s^{\rm cyl}\approx 2\pi R\left(1+\sqrt{\frac{k_r}{k_d}}\right)\quad.
\]
We observe that this is equivalent to a 2-dimensional flat surface with $2\pi R$ cells at the bottom. Notice that this is true for the asymptotics, but not in the transient, since the neutral competition between cells at the same level will give rise to different speeds towards monoclonality depending on the radius $R$ \cite{Lopez-Garcia:2010}. 

\subsubsection{Two-dimensional hemispheric geometry with radius R}

According to equation (\ref{eq:pckhemisphere}), the stochastic fluctuations will project over the surface from the bottom pole to the cell located at:
\[
z_s\approx 1+\sqrt{\frac{\pi}{2}\frac{k_r}{k_d}}\quad.
\]
The position of this cell will define a polar angle of:
\[
\varphi_s\approx\frac{1}{R}\left(1+\sqrt{\frac{\pi}{2}\frac{k_r}{k_d}}\right)\quad.
\]
Therefore, any cell located below this angle from the center will be considered a stem cell, and the stem cell number will be given by equation (\ref{eq:N_s_verygeneral}):
\begin{equation}
N_s^{2D}\approx 2\pi R^2\left[1-\cos\left\{\frac{1}{R}\left(1+\sqrt{\frac{\pi}{2}\frac{k_r}{k_d}}\right)\right\}\right]\quad.
\label{eq:N_sCrypt}
\end{equation}

\subsubsection{Spherical geometry}

In this case, the probability of lineage survival in terms of distance from the center of the sphere will read:
\[
p(c_n)\propto {\rm exp}\left\{-\frac{3k_d}{k_r}n^2\right\}\quad.
\]
The variance is expected to be: 
\[
\sigma\sim\sqrt{\frac{k_r}{3k_d}}\quad.
\]
In consequence, the effective stem cell number in a spherical system under the SCB dynamics taking place along the spherical coordinate can be approximated, according to equation (\ref{eq:N_s_verygeneral}), to:
\[
N_s^{\rm sphere}\approx \frac{4}{3}\pi\left(1+\sqrt{\frac{k_r}{3k_d}}\right)^{3} \quad.
\]

\subsection{A comment on tissue density, geometry and dynamical parameters}

Two main assumptions underlie the previous computations. The first one is that the density of the tissue is considered constant, so that we can normalize un-ambiguously all lengths/surfaces/volumes by the characteristic cell size. If cell-cell dispersion occurs via cell-cell intercalations, then we find that density changes do not change the dynamics, as the dynamics is fully independent of cell size. However, if dispersal occurs as in kidney with extrusion and random cell motility, then cells randomly moving over the same distance would intercalate over "more" cells in a high density setting (introducing a length scale measured in absolute length units and not in units of cell size). Thus, density would increase stem cell number from a purely dynamical effect. More quantitatively, if $k_r$ is given in terms of distance the cell moved in a time unit --instead of cell-cell intercalations--, then if the density $f$ of the tissue is $\lambda f$, with $\lambda >0$, that means that there will be $\lambda$ cells more for length/surface/volume unit. In that context, equation (\ref{eq:N_s_verygeneral}) reads:
\begin{equation}
N_s\approx \int . . .\int_0^{\lambda\sigma+1}\sqrt{g}dx_1. . . dx_{n-1}d{z}\quad,\nonumber
\end{equation}
implying in the case e.g., of a hemispheric geometry, that:
\[
N_s^{2D}\approx 2\pi R^2\left[1-\cos\left\{\frac{1}{R}\left(1+\sqrt{\lambda\frac{\pi}{2}\frac{k_r}{k_d}}\right)\right\}\right]\quad.
\]
To have an idea of what does it mean quantitatively, if we have a radius of $R=2$, as in the crypt, and $K_d=1$, $k_r=1$ in given distance units, $N_s^{2D}\approx 11$ cells. If the density doubles, i.e., $\lambda=2$, then, according to the above equation $N_s^{2D}\approx 23$.

A second assumption is that we have considered the tissue geometry (for instance the length and width of mammary/kidney tips) to be constant. If geometry changes during the growth process, for example enlarging the radius of the bud, then the stem cell number would also  change. The interplay between geometry and growing dynamics adds another layer of complexity to the SCB dynamics that goes beyond the scope of our current approach, and would be an interesting extension for the future.

\section{Noise in the stochastic conveyor-belt from "tectonic" epithelial movements}
\begin{figure}[h]
\begin{center}
\includegraphics[width= 8.9cm]{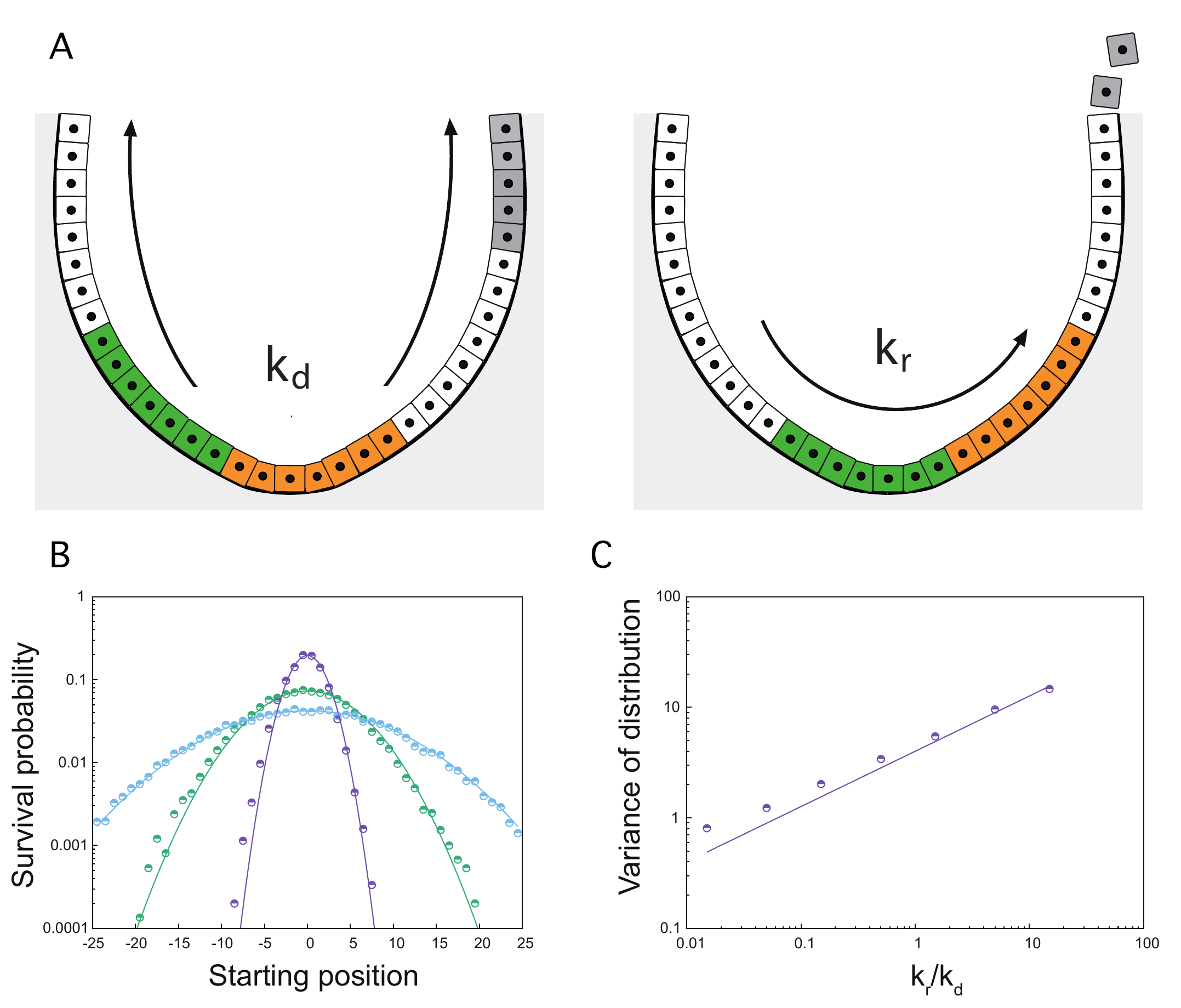}
\caption{A/ Schematic of the SCB dynamics where the source of stochasticity comes from  "tectonic" movements: cell division can occur for every cell, which pushes all cells above, but repositioning relative to the bottom of the crypt/tip occur via global movements of the layer, at rate $k_r$ (two clones shown competing before and after a movement). B/ Computational predictions of the 1-dimensional SCB dynamics in the presence of tectonic movements, with increasing rates $k_r$ (purple to blue), in terms of survival probability as a function of starting position of the clone. All curves are very well fitted by normal distributions, as expected by our model. C/ Variance of survival probability as a function of starting position (i.e. functional stem cell number) as a function of the tectonic movements rates normalized by division rate $k_r/k_d$ (dots), and theoretical prediction ($\sqrt{k_r/k_d}$, continuous line) from the SCB dynamics, showing that the overall behaviour of the system can be mapped to the one found with cell-cell random stochastic intercalations.}
\label{fig:Tectonic}
\end{center}
\end{figure}

In this section, we explore an alternative source for noise in determining the number of functional stem cells, i.e. the possibility of global rearrangements of the epithelium relative to the optimal position (bottom of the crypt/edge of the tip --see Fig. (\ref{fig:Tectonic}a) of this appendixfor a sketch. This is motivated by experiments in intestinal morphogenesis or mammary morphogenesis, where global three-dimensional bending of the epithelial modifies the location of the niche (see main text). 

Importantly, performing full stochastic simulations of this process in one-dimensions revealed a strikingly similar paradigm compared to the version of the model introduced in the main text, with survival probabilities decaying as normal distributions away from the central, optimal position for survival --see Fig. (\ref{fig:CryptGeom}b)  of this appendix. Moreover, the variance of these probabilities, which define the number of functional stem cells, also scale as $k_r/k_d$ as expected in the model from the main text --Fig. (\ref{fig:CryptGeom}c) of this appendix. 

This confirms that such tectonic movements can also be described in our coarse-grained model, simply renormalizing in long-term dynamics the intensity of the noise term $k_r$ in the system (although one would expect tectonic movements to significantly change the short-term dynamics). Interestingly, this allows for the system to be "noisy", i.e. many functional stem cells to contribute to the long-term dynamics, without any clonal dispersion, showing that one must be careful in equating the two directly. This would in particular be relevant for the dynamics of intestinal crypts, where cells away from starting position 0 have been shown experimentally to still contribute long-term (see Fig. 2 of the main text), but where little clonal fragmentation was observed (raising the possibility that such tectonic collective movements could occur to reposition cells towards/away from the best location).

\section{Numerical simulations and parameter estimation}
\begin{figure*}[h]
\begin{center}
\includegraphics[width= 17cm]{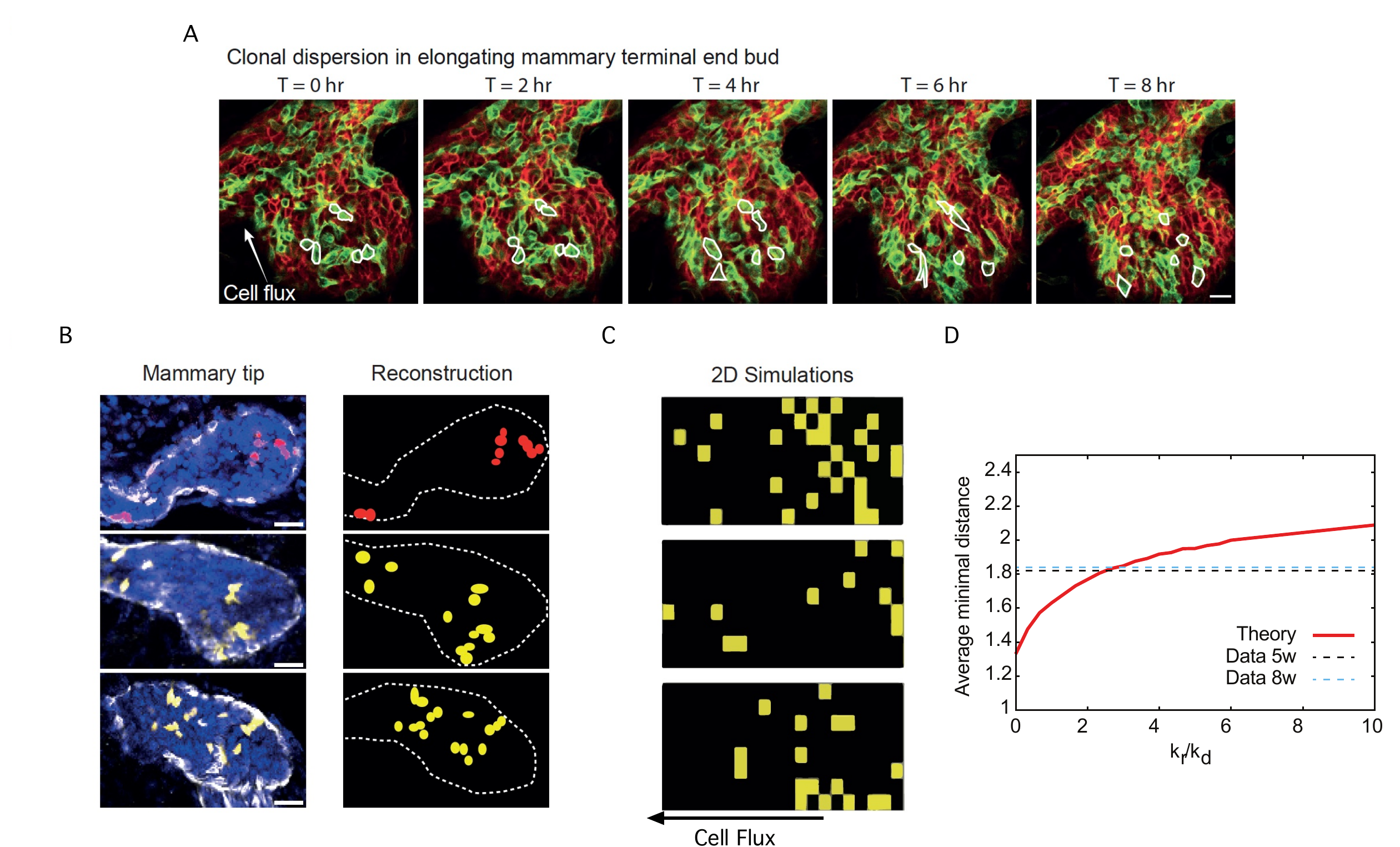}
\caption{Inferring $k_r$ from clone dispersion in mammary gland development A/
Intravital microscopy images of a developing terminal end bud followed over multiple hours showing extensive cell rearrangements leading to clonal dispersion. Some cells are highlighted with white lines to illustrate the random cell movements within the terminal end bud. Scale bar represents 10 $\mu$m. B/ 
Confocal images of terminal end buds stained with keratin 14 (white) to label the basal cells, and DAP (blue) to label the nuclei. Cells from the same lineage are marked in red or yellow. The reconstruction of the mammary tips is shown in the right panel, from which the minimal distance between cells from the same lineage was inferred. Scale bar represents 25 $\mu$m. 
 C-D/ Numerical simulations of clonal dispersion (three different outputs all with  $k_r/k_d = 3$, panel C). The theoretical average minimal distance between cells in a clone increase in $k_r/k_d$ (red line).  The value where the expected nearest neighbour distance cross the real data (5w and 8w, dotted lines) is the estimated $k_r/k_d \approx 3$. }
\label{fig:Fragmentation}
\end{center}
\end{figure*}

\subsection{1-dimensional simulation} A one-dimensional array of 20 cells was initialised, where every cell was given an index $x=\{0,1,...,19\}$ to identify their starting position in the crypt, corresponding to their lineage. 0 is the most advantageous position at the bottom of the crypt, and 19 is at the top where it will be removed from the crypt by any single division event below. The simulation parameters were $k_d$, the probability that a cell divides, and $k_r$, the probability that a cell switches positions with its neighbour in either direction. During the simulation, cells were chosen at random with equal probability to decide whether to divide or change neighbors. The simulation was terminated when the array was colonized by a single cell lineage --we denote this as a "win" by that lineage.

The probability of survival of a given lineage, $p(c_n)$ was calculated as the number of wins divided by the number of simulation runs. The simulation was repeated 2000 times. For every $\frac{k_r}{k_d}$, $p(c_n)$ was plotted as a function of the starting position $n$, and the data was fitted with the following function:
\begin{equation}
p(c_n) = \frac{1}{\sigma \sqrt{2 \pi}} {\rm exp} \Big\{ -\frac{n^2}{2\sigma ^2} \Big\}  \quad.
\end{equation}
Consistently with the theoretical predictions given in equation (\ref{eq:p(c_n)=}), we find that:
\[
\sigma\sim \sqrt{\frac{k_r}{k_d}}\quad.
\]

\subsection{2-dimensional simulation} 
A two-dimensional array of size 8x20 cells was initialised, where every cell was given an index $x=\{0,1,...,19\}$ to identify their starting row in the crypt, corresponding to their lineage. 0 is the most advantageous row and 20 is the upper boundary after which cells are removed. Note that this boundary condition is largely irrelevant for the results because cells at these rows have vanishing chance to contribute to a winning lineage. Periodic boundary conditions were applied to every row (cylindrical geometry). This simulation uses the same parameters $k_d$ and $k_r$, but now with a maximum of 4 possible neighbours for intercalation. The simulation was repeated 2000 times, and plotting the probability of survival of a given lineage shows that it fits to:
\begin{equation}
p_{2D}(c_n) = \frac{1}{\sigma \sqrt{2 \pi}} {\rm exp}\Big\{ -\frac{n^2}{2\sigma ^2} \Big\}  \quad.
\end{equation}	
In agreement with the theoretical predictions given in equation (\ref{eq:p(c_n)=}), we find that:
\[
\sigma\sim \sqrt{\frac{2k_r}{k_d}}\quad.
\]
Both fits for intestinal homeostatic crypts and developing kidney shown in the main text were performed using this 2-dimensional simulation. In both cases, in order to build confidence intervals, in a non-parametric way, for our predictions of clonal survival as a function of position, we simulated 2000 times the  number $N$ of labelling events as in the data ($N=45$ for intestine, $N=24$ for kidney), and calculated the mean survival probability, as well as the $68\%$ confidence interval around this prediction (i.e. 1 standard deviation around the mean). In both cases, we found that all of the experimental datapoint was contained within this interval. Finally, although on the long-term, survival probabilities converge towards a steady-state universal Gaussian distribution, the live-imaging datasets were experimentally acquired on finite timescales, requiring simulations to examine the dynamics of clonal conversion. We thus state below for each organ the duration of the simulations $T$ (rescaled by the cell division rate $k_d$).

For the intestinal crypts, clones were initialized only at positions 0,1,2,3 and 4, to match with the experimental set-up where clones were only traced from Lgr5+ stem cells. Moreover, we defined a clone as "lost" in the system as soon as it didn't have any cells in this compartment (positions 0-4), again to match the way that the experimental data was recorded in Ref. \cite{Ristma:2014}. One should note that this assumption is expected to be largely irrelevant for our findings, given the results of Section 5: $k_r/k_d$ in this system is small enough that $N_s < 5$, so that the probability for clones to come back in the Lgr5+ region after having left it during the early phase of clonal competition is vanishingly small.  Plots in the main figure for intestinal crypts used the following parameters: $k_r/k_d=1$ and a runtime of $T k_d = 0.5$ (we note that the latter is relatively small, corresponding experimentally to a bit less than one full cell division in 3 days in Ref. \cite{Ristma:2014}, which could be linked to the method of intravital imaging (typical timescales reported in intestine are 1-2 days).

For the kidney tips, clones were initialized evenly in positions 0-10, which was also the definition for the compartment of clonal survival. Plots in the main figure used the following parameters: $k_r/k_d=16$ (see main text for details on this parameter estimation) and a runtime of $T k_d = 2$ (which is the typical average number of divisions seen in the experimental dataset during the time course, see Ref. \cite{Riccio:2016}). The experimental data of Ref. \cite{Riccio:2016} assigns to cells a starting position on a 10x10 grid, and notes whether a clone still remains in the tip at the end of the observation period. As position $(10,10)$ was the edge of the tip, we calculated the euclidian distance of all coordinates $(i,j)$ from position $(10,10)$, which is the starting distance reported in Fig. 3 of the main text. One should note that because of the longer runtime of experiments, predictions in kidney are much-closer to their steady-state universal form.

\subsection{2-dimensional simulation and parameter estimation for fixed mammary gland samples} 

Finally, we also used these 2-dimensional simulations to fit the clonal fragmentation seen in mammary gland (where no long-term live-imaging was possible to follow clonal survival). To match experiments where cells were labelled in mouse of 3 weeks and collected at 5 weeks, and where the typical cell division rate is $16$ hours, we used $T k_d=20$, although we found that the predictions were largely insensitive to this timing. 

Given the low dose of clonal induction, all labelled cells in a tips of the same lineage (basal or luminal) can be considered to belong to a single clone. We thus experimentally measured for each labelled cell the distance to the closest labelled cell in a given tip, and built probability distributions for these nearest cell-cell distance across tips and mice (see details in section  6C of this appendix) . Note that this was done in 2D projections, so that this experimental distance is a 2D approximation of the real 3D dispersion.
We then performed the same computation for numerical simulations, for different values of $k_r/k_d$. As expected, for low values of $k_r/k_d$, cells are always close neighbours, so that the average cell-cell distance increased monotonously with $k_r/k_d$ (plotted in Fig. S6D). Comparing this theoretical prediction to the experimentally measured average minimal distance then allowed us to estimate $k_r/k_d \approx 3$, although this remains a rough estimate, given the uncertainty in the measurements and the approximations made in estimating 2D distances. However, as shown in Fig. 4 of the main text, the full numerical probability distributions for the nearest cell-cell distance with $k_r/k_d = 3$ provided a satisfactory fit for the experimental data (both 5w and 8w), strengthening the approach. We also show in Fig. S6C different examples of 2D stochastic simulations for the output of the model with $k_r/k_d=3$ (same parameter as in Fig. 4 of the main text) to give a better intuition of the variability of the clonal dispersion process.

\section{Experimental procedures}

\subsection{Mice}
All mice were females from a mixed background, housed under standard laboratory conditions, and received food and water ad libitum. All experiments were performed in accordance with the Animal Welfare Committee of the Royal Netherlands Academy of Arts and Sciences, The Netherlands.

\subsection{Intravital microscopy}
R26-CreERT2;R26-mTmG female mice were IP injected with 0.2mg/25gTamoxifen diluted in sunflower oil (Sigma) at 3 weeks of age to induce sporadic recombination in the developing mammary gland. At 5 weeks of age, a mammary imaging window was implanted near the 4th and 5th mammary gland --for details, see \cite{Alieva:2014}. Mice were anesthetized using isoflurane (1.5\% isoflurane/medical air mixture) and placed in a facemask with a custom designed imaging box. Imaging was performed on an inverted Leica SP8 multiphoton microscope with a chameleon Vision-S (Coherent Inc., Santa Clare, CA, www.coherent.com), equipped with four HyD detectors: HyD1 (<455nm), HyD2 (455--490nm), HyD3 (500--550nm) and HyD4 (560--650nm). Collagen I (second harmonic generation) was excited with a wavelength of 860nm and detected with HyD1, GFP and Tomato were excited with a wavelength of 960nm and detected with HyD3 and HyD4. Mammary gland tips were imaged at an interval of 20-30 minutes using a Z-step size of 3$\mu$m over a minimum period of 8 hours. All images were acquired with a 25$\times$ (HCX IRAPO N.A. 0.95 WD 2.5mm) water objective. 

\subsection{Quantitative image analysis}
Clonal dispersion in the developing mammary tips was measured in lineage traced whole mount glands from R26-CreERt2;R26-Confetti mice as previously described \cite{Scheele:2017} (data contained both re-analysis of glands from this paper, as well as ones from the same experiments which hadn't been analyzed before). In brief, R26-CreERt2;R26-Confetti female mice were injected at 3 weeks of age with 0.2mg/25g Tamoxifen to achieve clonal density labelling (<1 cell per tip). Lineage traced mice were sacrificed at mid-puberty (5 weeks of age) or at the end of puberty (8 weeks). Mammary glands were dissected, fixed in periodate-lysing-paraformaldehyde (PLP) buffer (1\% paraformaldehyde (PFA, Electron Microscopy Science), 0.01M sodium periodate, 0.075M L-lysine and 0.0375M P-buffer (0.081M Na2HPO4 and 0.019M NaH2PO4) (pH 7.4)) for 2 hours at room temperature (RT), and incubated for 2 hours in blocking buffer containing 1\% bovine serum albumin (Roche Diagnostics), 5\% normal goat serum (Monosan) and 0.8\% Triton X-100 (Sigma-Aldrich) in PBS. Subsequently, glands were incubated with primary antibodies anti-K14 (rabbit, Covance, PRB155P, 1:700) or anti-E-cadherin (rat, eBioscience, 14-3249-82, 1:700), and secondary antibodies goat anti-rabbit or goat anti-rat, both conjugated to Alexa-647 (Life Technologies, A21245 and A21247 respectively, 1:400). Mammary glands were mounted on a microscopy slide with Vectashield hard set (H-1400, Vector Laboratories), and imaging was performed using a Leica TCS SP5 confocal microscope, equipped with a 405nm laser, an argon laser, a DPSS 561nm laser and a HeNe 633nm laser. All images were acquired with a 20x (HCX IRAPO N.A. 0.70 WD 0.5mm) dry objective using a Z-step size of 5$\mu$m (total Z-stack around 200$\mu$m). 3-dimensional tile scan images of whole-mount mammary glands were used to manually reconstruct the tips. Labelled confetti cells were annotated in the schematic outline of the tips including information on the confetti colour for the mammary glands (GFP=green, YFP=yellow, RPF=red and CFP=cyan). The length and the width of the tips were measured, and the coordinates of each labelled confetti cell in the tip were determined. The coordinates were used to calculate the position of each labelled cell within the ductal tip, as well as the minimal distance to the nearest neighboring cells with the same confetti color (as a measure of the clonal dispersion within the stem cell zone).








%

\end{document}